\documentclass[12pt]{article}
\usepackage[yyyymmdd,hhmmss]{datetime}
\usepackage{graphicx}
\usepackage{pdfpages}
\usepackage{hyperref}
\usepackage{amssymb}
\usepackage{color}
\usepackage{ulem}
\usepackage{soul}

\def\text#1{\mbox{#1}}
\newcommand{\ie}{{\it{i.e.}}}
\newcommand{\eg}{{\it{e.g.}}}
\newcommand{\etc}{{\it{etc.}}}
\newcommand{\be}{\begin{equation}}
\newcommand{\ee}{\end{equation}}
\newcommand{\beq}{\begin{eqnarray}}
\newcommand{\eeq}{\end{eqnarray}}

\makeatletter
\@addtoreset{equation}{section}

\makeatletter
\renewcommand\section{\@startsection {section}{1}{\z@}%
                                   {-3.5ex \@plus -1ex \@minus -.2ex}
                                   {2.3ex \@plus.2ex}%
                                   {\normalfont\large\bfseries}}
\renewcommand\subsection{\@startsection{subsection}{2}{\z@}%
                                     {-3.25ex\@plus -1ex \@minus -.2ex}%
                                     {1.5ex \@plus .2ex}%
                                     {\normalfont\bfseries}}

\def\ttau{{\cal T}}

\begin{document}

\begin{center}

{\bf \Large
Thermal partition function of $J_3 \bar J_3$ deformed $AdS_3$
}

\vspace{1cm}

Soumangsu Chakraborty$^1$, Amit Giveon$^2$,  Akikazu Hashimoto$^3$

\vspace{1cm}

{}$^1$ Universit\'e Paris-Saclay, CNRS, CEA, \\Institut de Physique Th\'eorique, 91191 Gif-sur-Yvette, France

{}$^2$Racah Institute of Physics, The Hebrew University \\
Jerusalem 91904, Israel

{}$^3$ Department of Physics, University of Wisconsin \\ Madison, WI 53706, USA

\end{center}

{\it E-mail:} \href{mailto:soumangsuchakraborty@gmail.com}{soumangsuchakraborty@gmail.com}, \href{mailto:giveon@mail.huji.ac.il}{giveon@mail.huji.ac.il},
\href{mailto:hashimoto@wisc.edu}{aki@physics.wisc.edu}

\vspace{1cm}

\begin{abstract}
\noindent We derive a compact formula for the one-loop, bosonic string partition function of Euclideanized $J_3 \bar J_3$ deformed $AdS_3$ with periodic Euclidean time as an integral transform of the partition function of the undeformed Euclideanized $AdS_3$. Such a deformation is interpretable as an irrelevant ``single-trace $T \bar T$ deformation" of the boundary. We will do this by first establishing a formal procedure to compute a worldsheet torus zero point function for an exactly marginal $J \bar J$ deformation of a sigma model with $U(1)_L \times U(1)_R$ global symmetry.  We then describe how this procedure is implemented on $SL(2,R)$ sigma model and its Euclidean continuation. Finally, we describe the embedding of the deformed $SL(2,R)$ torus amplitude into critical string theory and interpret the result as the leading perturbative contribution to the thermal partition function of the deformed theory.

\end{abstract}


\newpage

\tableofcontents

\section{Introduction}

Anti-de-sitter space in three dimensions is an important background in string theory. Such a background is related to conformal field theories in two dimensions via holography  (see  \cite{Aharony:1999ti} for a review). When the $AdS_3$ is supported by a NSNS three-form flux field, in the weak coupling limit, it admits a worldsheet description in terms of a $SL(2,R)$ WZW sigma model at level $k$ combined with some CFT ${\cal N}$ \cite{Maldacena:2000hw}. For the particular case of superstrings on $AdS_3\times S^3\times T^4/K3$ with NSNS three-form flux, the background can be thought of as obtained from a stack of parallel $p$ F1-strings on $k$ NS5-branes in its near-horizon limit.

It was realized right away that $AdS_3$ background contains wound states that can reach the boundary of $AdS_3$ with finite energy \cite{Maldacena:1998uz,Seiberg:1999xz}. Such states are referred to as ``long strings''. The spectrum of free strings in this background was analyzed in \cite{Maldacena:2000hw} who showed that the states consist of discrete and continuum representations of $SL(2,R)$. The long strings correspond to the continuum states. The discrete states are often termed in the literature as the ``short strings".  The Euclidean continuation of $AdS_3$ is also called $H_3 = SL(2,C)/SU(2)$.  String theory on $H_3 \times {\cal N}$ with  time compactified  to period $\beta$ is also referred to as string theory on ``thermal $AdS_3$'' or ``Euclidean BTZ'' times ``internal space". The one-loop string amplitude in this background was computed in \cite{Maldacena:2000kv}, and is interpretable as the thermal Boltzmann sum of the states identified in  \cite{Maldacena:2000hw} along the lines of \cite{Polchinski:1985zf}.

In \cite{Giveon:2017nie,Giveon:2017myj}, a solvable deformation of the $SL(2,R)$ sigma model by a marginal deformation $J_- \bar J_-$ was considered.  This deformation gives rise to a background that interpolates between a linear dilaton background at a large radius and the zero mass BTZ solution at a small radius. Holographically, this deformation is interpretable as an irrelevant deformation giving rise to non-local physics of the little string theory (LST) in the UV, flowing to $1+1$ CFT in the IR.  The full spacetime CFT dual to perturbative string theory in $AdS_3$ is not known in general but it is believed that the long strings sector of the theory has a symmetric product structure. It is also possible to show  \cite{Giveon:2017myj,Hashimoto:2019wct} that the target space theory of a single long probe string in the deformed background does appear to deform like $T \bar T$  of \cite{Smirnov:2016lqw,Cavaglia:2016oda}. Because of this, one often refers to the holographic duality in string theory  on the  background in
 \cite{Giveon:2017nie,Giveon:2017myj} as   ``single-trace $T \bar T$ holography". \footnote{For the present status of single-trace $T \bar T$ holography see \eg\ \cite{Chakraborty:2023mzc,Chakraborty:2023zdd,Chakraborty:2024ugc}. }

In this article, we consider a different but closely related solvable deformation of $SL(2,R)$ sigma model by a marginal operator $J_3 \bar J_3$  and its embedding into critical string theory.  The resulting string theory background interpolates between a linear dilaton background at a large radius and a global $AdS_3$  at a small radius. The topology of this background is such that the spatial circle at a fixed radius is contractable near the origin. From the holographic perspective, this construction can be regarded as the ``single-trace $T \bar T$ deformation" of string theory in global $AdS_3$. Here the boundary field theory flows from LST at short distances to a CFT$_2$ in its  $SL(2,R)$ invariant ground state (the NS ground state in the SUSY case) in the IR. As in the case of  $J_-\bar{J}_-$ deformation of massless BTZ, $J_3\bar{J}_3$ deformation of global $AdS_3$  acts as the $T\bar{T}$ deformation of the spacetime theory of a single long string while the full long string sector takes the form of a symmetric product of $T\bar{T}$ deformed long string CFT. The discrete states however do not have a symmetric product structure. In this article, we compute the dimensions of worldsheet operators of $J_3 \bar J_3$ deformed $SL(2,R)$ WZW model.  Embedding it in critical string theory we compute the perturbative spectrum of both the discrete and the continuous states of the spacetime theory to further support the above-mentioned symmetric product structure of the long strings.

The main goal of this article is to study the spectrum and the one-loop vacuum amplitude  (interpretable as the thermal partition function) for the $J_3 \bar J_3$ deformed $AdS_3$, and its Euclidean continuation $H_3$,  embedded into critical string theory. It is straightforward to compute the spectrum by formulating the deformed theory as a gauged sigma model. The partition function for  $H_3$ with Euclideanized time coordinate compactified with period $\beta$ turns out to be a little involved. We will show that this quantity, for the deformed theory, can be computed by utilizing the folding/unfolding trick which was used in the analysis of \cite{Polchinski:1985zf} and an integral kernel formula for computing the partition function deformed by the bilinear operator of the form $J \bar J$. That a bilinear operator deformation leads to an integral transform of the partition function was found in the context of double-trace deformation of matrix models in \cite{Gubser:2002vv}. Similar results were obtained for $T \bar T$ and $J \bar T$ deformation in \cite{Hashimoto:2019wct}. We will derive an integral kernel for the $J_3 \bar J_3$ of the $SL(2,R)$ sigma model which reproduces the deformed spectrum. We will then use this kernel formula and the folding/unfolding trick to compute the one-loop partition function for the $J_3 \bar J_3$ deformed $H_3$ target space compactified with period $\beta$ and twist $\mu$,  and provide various tests by embedding the result into string theory and computing the one-loop zero-point string amplitude.

``Single-trace $T\bar{T}$ deformation" of string theory in global $AdS_3$ was considered previously in \cite{Apolo:2019zai} using TsT techniques.  The TsT deformation appears to be related to $J \bar J$ deformation, and as such,  our analysis of the non-linear sigma model background associated with the deformed theory and spectrum of the continuum states of the spacetime theory partially overlaps with the work of \cite{Apolo:2019zai}. In this work, we do a thorough analysis of the discrete as well as the continuum states by analyzing the $J_3 \bar J_3$ deformed worldsheet sigma model systematically.
 We also discuss how the additive constant component of the NSNS $B$-field, discussed in \cite{Chakraborty:2024ugc}, affects the spectrum and the partition function. From the thermal partition function of the spacetime theory, we derive a closed-form expression of the density of states of the spacetime long strings and show that it smoothly interpolates between the density of long string states in $AdS_3$ computed in \cite{Maldacena:2000kv} in the IR to that of the cigar in the UV \cite{Hanany:2002ev}.

The organization of this paper is as follows. In section \ref{sec:scalar}, we analyze, in detail, the $J_x \bar J_x \sim \partial x \bar \partial x$ deformation of a free, compact scalar field $x$ with period $x = x + \beta$. This is a trivial, solvable system where the effect of the deformation simply changes the proper radius of the compact scalar. Despite its triviality, we will analyze it using a chain of arguments that may seem unnecessarily complicated at first sight. The point of section \ref{sec:scalar} is to establish a template for the analysis of the more sophisticated $H_3$ and $SL(2,R)$ case. We will introduce concepts such as the gauged sigma model formulation of the marginal deformation,  chemical potentials, folding/unfolding trick, and the kernel in the scalar context first. Then, in section \ref{sec:sl2r}, we will consider the $H_3$ model. We will provide a precise mathematical formulation of the partition function we are interested in computing in section \ref{sec:problem}, and then trace the arguments outlined in section \ref{sec:scalar}.  In section \ref{sec:answer}, we provide the solution to the problem posed in section \ref{sec:problem}. The solution involves an integral kernel. The fact that the action of the kernel on a Boltzmann term reproduces the spectrum of the deformed theory serves as a consistency check of our solution. In section \ref{sec:embed}, we describe the target space physics encoded in the string one-loop partition function obtained by embedding $H_3$ into critical string theory on $H_3 \times {\cal N}$. We describe the status of spectral decomposition where one expresses the worldsheet partition function as a trace of $\exp(2 \pi i(\tau L_0 - \bar \tau \bar L_0))$ in section \ref{sec:trace}.  While such a decomposition is standard in conformal field theories, the non-compactness of $H_3$ arising from the existence of long strings makes computation and manipulation of this quantity extremely subtle. To highlight the thermal interpretation of the one-loop amplitude on $H_3 \times {\cal N}$ and its deformation manifest, the spectral decomposition is extremely convenient. In section \ref{secjjstemded}, we show that the spectrally decomposed form can be used to compute the density of states of the long strings in the deformed spacetime theory with all the desired properties.  A flow chart illustrating the logical structure of these arguments is provided in figure \ref{fige} in the discussion section.  In appendix \ref{sec:appa}, we describe some technical aspects of performing the path integral of gauged $SL(2,R)$ sigma model.

\section{$J_x\bar J_x$ deformation of a free scalar boson theory \label{sec:scalar}}

In this section, we will analyze vacuum torus amplitude for the $J_x \bar J_x$ deformed  free scalar theory
\be S = {1 \over 2 \pi} \int dz_1 \, dz_2 \,  \partial x \bar \partial x ~, \label{xaction}\ee
where we impose a periodicity
\be x \sim x + \beta~, \ee
and\footnote{These currents are defined with the normalization that $\epsilon(J_x + \bar J_x)$ corresponds to the shift $x \rightarrow x + 4 \pi \epsilon$. See (2.3.12) and (2.3.13) of \cite{Polchinski:1998rq}.}
\be J_x(z) =i \partial x~, \qquad \bar J_x(\bar z) =i \bar \partial x~,  \ee
so that
\be J_x(z) J_x(0) \sim  {1 \over z^2} \ .\label{jxjxope} \ee
The coordinate radius, $R_x$, of the compact scalar field $x$ is related to $\beta$ via the equation  $\beta=2\pi R_x$.

This is a trivial deformation of a trivial theory.  We will nonetheless analyze this deformation extremely thoroughly. In fact, we will describe {\it three} different approaches to analyze this problem. The point is that the more seemingly roundabout approach is more easily adaptable to the analysis of the $SL(2,R)$ case and its Euclidean continuation.

 Since the free boson theory is extremely simple, the approach we take will likely come across as unnecessarily complex. The point, of course, is that this slightly expanded approach will apply naturally when we consider the more complicated $SL(2,R)$ theory. The free boson theory is a lot more convenient for establishing some of the subtle steps we will be implementing in the later sections.

Let us start by establishing the convention
\be \int dz_1 \, dz_2  = (2\pi)^2 \tau_2 ~, \ee
and
\be z = z_1 + i z_2 \sim z + 2 \pi \sim z + 2 \pi \tau~. \label{z1z2} \ee
Some care is needed in comparing to literature \eg\ (2.1.7) of \cite{Polchinski:1998rq} where
\be d^2 z   = i dz\, d \bar z\ = 2 d z_1\, d z_2 ~. \ee

We will implement the $J_x \bar J_x$ deformation by adding two auxiliary fields: (a) a timelike, non-compact field $t(z,\bar z)$ and (b) a compact spacelike field $y(z, \bar z)$ with periodicity $y\sim y+2\pi R_y$, and gauging the combined system as follows. Consider a set of holomorphic and anti-holomorphic null current
\be J  = l_1 J_x + l_2 J_t + l_3 J_y~, \qquad \bar J = l_1 \bar J_x + l_2 \bar J_t - l_3 \bar J_y~, \label{currgauge} \ee
where the currents $J_t = i \partial t$  and $J_y = i \partial y$ with the following normalizations
\be J_t(z) J_t(0) \sim -{ 1 \over  z^2}~, \qquad J_y(z) J_y(0) \sim { 1 \over  z^2} \ . \ee
By setting
\be  l_1^2 - l_2^2 + l_3^2 = 0~, \label{nullcond}\ee
one can make the gauge current, $J(z)$ and $\bar J(z)$ null. In other words, the OPE
\be J(z) J(0) \sim \mbox{regular} \ . \ee
Since $J$ is null, we can (arbitrarily) set the overall normalization of $J$ by setting
\be l_3 = 1 \ .\label{l3=1} \ee
If so, the null condition (\ref{nullcond}) implies $l_2 = \sqrt{1 +  l_1^2}$ and so the space of charges $(l_1, l_2, l_3)$ is parameterized by a single quantity $l_1$.

Gauging the system of $(x,t,y)$ fields independently with respect to $J(z)$ and $\bar J(z)$ leads to an action of the form
\be S =    \int d z_1 \, d z_2\, {\cal L} \ , \ee
where
\beq 2 \pi {\cal L} & =&  \left( \partial x \bar \partial x + 2 l_1 B\bar \partial x + 2 l_1 \bar A  \partial x + 2 l_1^2 B \bar A \right)\cr
&& -  ( \partial t \bar \partial t - 2l_2 B\bar \partial t -2 l_2 \bar A  \partial t + 2 l_2^2 B \bar A ) \cr
&& +  \partial y \bar \partial y - 2 B\bar \partial y + 2  \bar A  \partial y - 2  B \bar A~,  \label{sbasicaction2}
\eeq
and is invariant (up to total derivatives) under the transformation   \footnote{At this stage we regard $A_\alpha,B_\alpha$ as independent gauge fields.}
\beq x(z, \bar z) & \rightarrow & x(z, \bar z) + l_1 \Lambda(z, \bar z) + l_1 M(z, \bar z)~, \cr
t(z, \bar z) & \rightarrow& t(z, \bar z) - l_2 \Lambda(z, \bar z) - l_2 M(z, \bar z)~, \cr
y(z, \bar z) & \rightarrow & y(z, \bar z) + \Lambda(z, \bar z) - M(z, \bar z) ~,\cr
A(z,\bar z) & \rightarrow & A(z,\bar z) - \partial \Lambda(z,\bar z) ~,\cr
\bar A(z, \bar z) & \rightarrow & \bar A(z, \bar z) - \bar \partial \Lambda(z, \bar z)~, \cr
B(z, \bar z) & \rightarrow &  B(z,\bar z) - \partial M(z, \bar z)~, \cr
\bar B(z, \bar z) & \rightarrow & \bar B(z, \bar z) - \bar \partial M(z, \bar z) \ . \label{sgauge}\eeq
The idea to gauge the left and right movers independently as is done in (\ref{sbasicaction2}) for field theories in 2 dimensions was first proposed in \cite{Bars:1991pt}  and has since been discussed \eg\ in  \cite{Ginsparg:1992af,Quella:2002fk,Israel:2003ry,Israel:2004ir,Martinec:2020gkv,Bufalini:2021ndn} and is frequently used in gauging WZW sigma models. We will refer to this type of gauging as ''asymmetric gauging'' to distinguish it from the more ordinary gaugings based on minimal coupling.  One can represent the content of this gauging in the coset form
\be {U(1)_x \times R_t \times U(1)_y \over U(1)_L \times U(1)_R} \ . \label{aaa}\ee
Invariance under large gauge transformation constrains the period of $y$ and the period of $x$ by relation $R_y = R_x/l_1$.
It is clear that if $l_1 = 0$, the gauging does not affect the $x(z,\bar z)$ field and corresponds to the undeformed limit.

In the rest of this section, we will analyze the vacuum partition function of this system on a torus with modular parameter $\tau$. This quantity,
\be Z^{def\ x}_{cft}(\tau, \bar \tau, \beta, l_1) \ ,  \ee
depends on $\tau_1$, $\tau_2$, $\beta$, and $l_1$. We will provide three separate analyses. They will all lead to the same conclusion but will highlight different features. Going through these three separate analyses will facilitate the process of understanding the $SL(2,R)$ model.

\subsection{Integrating out the auxiliary fields\label{sec11}}

In this subsection, we will compute $Z^{def\ x}(\tau, \bar \tau, \beta, l_1)$ by simply integrating out the auxiliary fields in (\ref{sbasicaction2}). There are some choices in the order of integrating out various fields. Let us first present one particular order that highlights the relation between null gauging and more ordinary gauging.

Let us start by imposing a partial gauge fixing condition
\be \partial \bar A - \bar \partial B = 0 \ . \label{partialgauge}\ee
We can then integrate out the $y$ field using the formula for the gauge field effective action of Polyakov and Wiegmann \cite{Polyakov:1984et,Polyakov:1988qz}
\beq \lefteqn{ \partial y \bar \partial y - 2 l_3 B\bar \partial y + 2 l_3 \bar A  \partial y - 2 l_3^2 B \bar A } \cr
&\rightarrow  & - l_3^2 B {\bar \partial \over \partial } B-{l_3^2}  \bar A {\partial \over \bar \partial} \bar A +  2 l_3^2 B \bar A - 2 l_3^2 B \bar A \cr
& = & - 2 l_3^2 \bar A B \cr
& = & 2 l_1^2 \bar A B - 2 l_2^2 \bar A B\ . \eeq
The residual gauge symmetry restricts
\be \Lambda(z,\bar z) = M(z,\bar z) ~,\ee
so we might as well call $B=A$. This will make the action read
\beq 2 \pi {\cal L} & =&    \left( \partial x \bar \partial x + 2 l_1 A\bar \partial x + 2 l_1 \bar A  \partial x + 4 l_1^2 A \bar A \right)\cr
&& -  ( \partial t \bar \partial t - 2l_2 A\bar \partial t -2 l_2 \bar A  \partial t + 4 l_2^2 A \bar A )~. \label{sbasicaction3}
\eeq
In this form, the $x$ and $t$ fields are minimally gauged as a linearly charged field. We can therefore represent this model as a timelike coset
\be\frac{ U(1)_x \times R_t}{ U(1)} ~, \label{bbb}\ee
and the gauge invariance is completely manifest.

We can now impose a gauge condition that $t=0$ and integrate out $A,\bar{A}$;
\be 2 \pi {\cal L}  =    \left(1 + l_1^2 \right) \partial x \bar \partial x = \partial x \bar \partial x- l_1^2 J_x^2  =  l_2^2  \partial x \bar \partial x~. \label{VV0ratio} \ee
This reproduces the expectation that $J_x \bar{J}_x$ deformation rescales the ``proper length'' of the target space parameterized by $x$.  \footnote{The deformation that decreases the size of $x$ is obtained by a vector gauging in the $x$ direction in (\ref{sbasicaction2}),(\ref{sbasicaction3}); this will give rise to  (\ref{VV0ratio}) with $l_1^2\to -l_1^2 $.}

We can also obtain (\ref{VV0ratio}) by imposing the gauge $t=y=0$ to (\ref{sgauge}) and integrating out $\bar A$ and $B$.  \footnote{As such we could have skipped the discussion between (\ref{partialgauge}) and (\ref{VV0ratio}), but we found it useful to review the relation between asymmetric gauging and ordinary gauging. Specifically, the coefficient of $l_1^2 A \bar A$ term in (\ref{sbasicaction3}) is 4 and allows for the completion of the square line by line, but the coefficient of $l_1^2 \bar A B$ in (\ref{sbasicaction2}) is 2 and the expression does not form a complete square. One can confirm by explicit analysis that (\ref{sbasicaction2}) is invariant with respect to (\ref{sgauge}), but the analysis between (\ref{partialgauge}) and (\ref{VV0ratio}) can be considered a quick consistency check of gauge invariance.}

Let us now complete the computation of
\be Z^{def\ x}_{cft}(\tau, \bar \tau, \beta, l_1) = \int [Dx] e^{-S} \ . \ee
This can easily be done by looking at (8.2.11) of \cite{Polchinski:1998rq} but substituting
\be \alpha' = 2  \times {1 \over  l_2^2} ~, \qquad 2 \pi R = \beta \ , \ee
where we account for the normalization of the  $x$ field in the action in our convention and in \cite{Polchinski:1998rq}. Upon Poisson resummation which takes (8.2.11) to (8.2.9), we arrive at
\be Z^{def\ x}_{cft}(\tau, \bar \tau, \beta, l_1)
= |\eta(\tau)|^2 \sum_{n,w} \exp\left[-\pi \tau_2 \left( {8 \pi^2 } {n^2 \over (l_2\beta)^2} + {1 \over 8 \pi^2}w^2 (l_2 \beta)^2- {1 \over 6} \right) + 2 \pi i \tau_1  n w \right]~. \label{zxdef} \ee
We see that this partition function satisfies a simple relation
\be Z^{def\ x}_{cft}(\tau, \bar \tau, \beta, l_1)  = Z^{x}_{cft}(\tau, \bar \tau,\tilde \beta=l_2 \beta)~, \label{l2scale} \ee
and is simply a reflection of the fact that a free periodic scalar $x \sim x + \beta$ with action (\ref{VV0ratio}) is the same as that of a free periodic scalar $\tilde x = l_2 x$ whose period is $\tilde x \sim \tilde x + \tilde \beta$ for $\tilde \beta = l_2 \beta$.

The expression (\ref{VV0ratio}) is also an explicit confirmation that the $U(1)_x \times R_t \times U(1)_y /( U(1)_L \times U(1)_R)$ asymmetrically gauged $(x,t,y)$  system is the deformation of free scalar action ${\cal L}$ given in (\ref{xaction}) by $J_x \bar J_x$ with coefficient $l_1^2/2\pi$.

\subsection{$J_x \bar J_x$ deformation as a kernel}

In this subsection, we will show that the expression for the partition function of the $J \bar J$ deformed model  (\ref{zxdef}) can be expressed in the form of a kernel acting on the partition function of the undeformed theory along the line similar to \cite{Hashimoto:2019wct}. This is achieved by evaluating the path integral of (\ref{sbasicaction2}) in a different choice of gauge and the order of integrating out the field.

Specifically, let us use the freedom of $\Lambda(z, \bar z)$ and $M(z,\bar z)$ to make all but the zero mode components of the gauge fields to zero, so that
\be \bar A(z,\bar z) = \bar A_0~, \qquad B(z, \bar z) = B_0 \ . \ee
We can then integrate out the $t$ and $y$ fields to arrive at
\beq 2 \pi {\cal L}& =&     \left(\partial x \bar \partial x + 2 l_1 B_0 \bar \partial x + 2 l_1 \bar A_0  \partial x + 2   l_1^2 B_0 \bar A_0 \right) \cr
&& -   2 l_2^2 B_0 \bar A_0  \cr
&& - 2 B_0 \bar A_0~, \eeq
which can also be written as
\beq  2 \pi {\cal L} & =&   \left( \partial x \bar \partial x + 2 l_1 B_0 \bar \partial x + 2 l_1 \bar A_0  \partial x + 4   l_1^2 B_0 \bar A_0 \right) \cr
&& -   4 l_2^2 B_0 \bar A_0\ .  \label{tyint2}
\eeq
An easy way to do these integrals is to use (A.9)--(A.14) of \cite{Hashimoto:2019wct} in the limit of large $r$.

Next, we wish to perform the path integral over the $x$ field in  (\ref{tyint2}) and present the result in the form of the trace of $\exp(2 \pi i (\tau L_0 - \bar \tau \bar L_0))$ and interpret the dependence on $\bar A_0$ and $B_0$. Once again, we can use the analysis in Appendix A of \cite{Hashimoto:2019wct} and define
\be Z_{inv}^x(B_0, \bar A_0) = \int [Dx] e^{-{1 \over 2 \pi}\int d z_1\,  d z_2 \left( \partial x \bar \partial x + 2 l_1 B_0 \bar \partial x + 2 l_1 \bar A_0 \partial x \right)}~. \label{zinv}
\ee
It is also convenient for later to define
\beq
Z_{asym}^x(B_0, \bar A_0) &=& \int [Dx] e^{-{1 \over 2 \pi} \int d z_1\,  d z_2 \left( \partial x \bar \partial x + 2 l_1 B_0 \bar \partial x + 2 l_1 \bar A_0 \partial x + 2   l_1^2 B_0 \bar A_0\right)}~, \cr
Z_{minimal}^x(B_0, \bar A_0) &=& \int [Dx] e^{-{1 \over 2 \pi} \int d z_1\,  d z_2\left( \partial x \bar \partial x + 2 l_1 B_0 \bar \partial x + 2 l_1 \bar A_0 \partial x + 4   l_1^2 B_0 \bar A_0\right)}  ~,
\eeq
which are related to $Z^x_{inv}(B_0, \bar A_0)$ as follows
\be Z_{inv}^x(B_0, \bar A_0) = Z_{asym}^x(B_0, \bar A_0) e^{-  l_1^2 B_0 \bar A_0/ \pi}= Z_{minimal}^x(B_0, \bar A_0) e^{-2  l_1^2 B_0 \bar A_0/\pi} \ .  \ee

If we now rescale
\be - {i \over \tau_2} \bar \xi = 2 l_1 B_0~, \qquad -{i \over \tau_2}  \xi = 2 l_1 \bar A_0~, \  \label{chemgauge}\ee
we can write (\ref{zinv}) as
\be Z_{inv}^x(\xi, \bar \xi) = \int [Dx] e^{-{1 \over 2 \pi}  \int d z_1\,  d z_2  \left( \partial x \bar \partial x - {i \over \tau_2}   \xi  \partial x -  {i \over \tau_2} \bar \xi  \bar \partial x \right)} ~.
\ee
This expression is recognizable as (A.14) of \cite{Hashimoto:2019wct}. Then, we can use (A.9) and (A.11) of \cite{Hashimoto:2019wct} and write
\be Z_{inv}^x(\xi, \bar \xi)=Z_{cft}^x(\xi, \bar \xi) e^{ \pi (\xi - \bar \xi)^2/ \tau_2} ~,\label{cftinv}\ee
where
\be Z^{x}_{cft}(\tau, \bar \tau, \xi, \bar \xi) =
 \sum_{states} \exp\left[2 \pi i (\tau (\Delta_0+N) - \bar \tau (\bar  \Delta_0+ \bar N)-{\tau_2 / 12}) + 2 \pi i \xi \mathfrak{j}_0 - 2 \pi i \bar \xi \bar \mathfrak{j}_0 \right]~, \label{Zxcft} \ee
where
\be \Delta_0 + \bar \Delta_0  = {4 \pi^2 } {n^2 \over \beta^2} + {1 \over 16 \pi^2}w^2 \beta^2  ~, \qquad \Delta_0 - \Delta_0 = n w + N - \bar N~, \ee
and
\beq\label{j0j0bar}
        \mathfrak{j}_0&=&\frac{n}{R_x}+\frac{wR_x}{2}~, \cr
        \bar \mathfrak{j}_0&=&\frac{n}{R_x}-\frac{wR_x}{2}~,
    \eeq
with  $R_x=\beta/(2 \pi)$ are the charges of the states with respect to the currents $2 \pi J_x$ and $2 \pi \bar J_x$ (See (2.3.14a) and (2.3.14b) of \cite{Polchinski:1998rq} regarding the origin of this factor of $2 \pi$.)
\be J_x(z) e^{i k x}(0,0)  = {k_L \over z} e^{i k x}(0,0)+ \ldots~, \qquad \bar J_x(\bar z) e^{i k x}(0,0)  = {k_R \over \bar z} e^{i k x}(0,0)+ \ldots \ ,  \ee
and $N,\bar{N}$ are the left and right moving oscillator numbers.

Now, let us go back to (\ref{tyint2}) and complete the computation
\be \fbox{\parbox{4.5in}{$$ Z^{def\ x}_{cft}(\tau,\bar{\tau}) = \int {d \xi\, d \bar \xi \over \tau_2} {l_2 \over l_1^2} e^{\pi (\xi - \bar \xi)^2/ \tau_2} e^{- 2 \pi \xi \bar \xi / \tau_2 l_1^2 }  Z^x_{cft} (\tau, \bar \tau, \xi, \bar \xi) ~.
$$}} \label{c1kernel2} \ee

We can recognize (\ref{c1kernel2}) as a kernel acting on the partition function of the undeformed partition function $Z^x_{cft} (\tau, \bar \tau, \xi, \bar \xi)$ where one integrates over the chemical potential $(\xi, \bar \xi)$. The factor of $l_1^2$ arises from the change of variable (\ref{chemgauge}). The factor of $\tau_2$ was adjusted to reproduce the correct $l_1 \rightarrow 0$ limit, and the factor of $l_2$ was adjusted so that the deformed partition function is naturally normalized. If one were to apply the kernel term by term on $Z^x_{cft}$ expressed in the form (\ref{Zxcft}), one finds
\be Z^{def\ x}_{cft}  (\tau, \bar \tau)= \sum_{states}   \exp\left[2 \pi i (\tau (\Delta+N) - \bar \tau (\bar \Delta+ \bar N)-\tau_2/12) \right] \ , \label{Zxdef2}\ee
where
\be \Delta + \bar \Delta  = {4 \pi^2} {n^2 \over (l_2 \beta)^2} + {1 \over 16 \pi^2}w^2 (l_2 \beta)^2 ~, \qquad \Delta - \bar \Delta = n w +N - \bar N \ , \label{Deltas} \ee
in agreement with (\ref{zxdef}).

\subsection{Analysis using folding and unfolding trick \label{sec:fold}}

In this subsection, we will describe another approach to compute the partition function of the $J_x \bar J_x$ deformed compact scalar theory. The approach we describe utilizes the folding/unfolding trick which was used extensively in  \cite{Maldacena:2000kv,Polchinski:1985zf} and is reviewed in \cite{Trapletti:2002uk}. One can efficiently generalize this approach to the case of $SL(2,R)$ sigma model.

Our starting point is (8.2.11) of \cite{Polchinski:1998rq} which we reproduce below
\be Z^x_{cft}(\tau, \bar \tau,\beta) = \sum_{m,w=-\infty}^\infty Z^x_{(m,w)}(\tau,\bar{\tau})=  {R_x \over \sqrt{\alpha' \tau_2}}| \eta(\tau)|^2 \sum_{m,w=-\infty}^\infty \exp\left( - {\pi R_x^2 | m - w \tau|^2 \over \alpha' \tau_2} \right) ~,\ee
where in our convention,
\be \alpha' = 2~, \qquad 2 \pi R_x = \beta \ . \ee
When summed over all $m$ and $w$, $Z^x_{cft}(\tau, \bar \tau)$ is modular invariant.

The folding/unfolding trick utilizes the fact that one can exchange the sum over $w$ by setting $w=0$ and summing over the orbit of $\Lambda \in SL(2,Z)/Z$ where $Z$ is generated by the $T$ element $\tau \rightarrow \tau +1$ of $SL(2,Z)$, and write
\be Z^x_{cft}(\tau, \bar \tau,\beta) =   Z^x_{(0,0)}(\tau, \bar \tau) +{R_x \over \sqrt{\alpha' \tau_2}}| \eta(\tau)|^2 \sum_{m=1}^\infty \sum_{\Lambda} \exp\left( - {\pi R_x^2 m^2 \over \alpha' (\Lambda \tau_2)} \right) ~.\ee
Like in \cite{Maldacena:2000kv}, the contribution from the $(0,0)$ sector is independent of $\beta$ and will be ignored in the remainder of our analysis.
The orbit of $\Lambda$ acting on the fundamental domain ${\cal F}$ spans the strip $-1/2 < Re(\tau) < 1/2$.  If we further define
\be Z^x_{(1,0)}(\tau, \bar \tau,\beta) =   {R_x \over \sqrt{\alpha' \tau_2}}| \eta(\tau)|^2  \exp\left( - {\pi R_x^2  \over \alpha'  \tau_2} \right)~, \ee
then we can write
\be Z^x_{cft}(\tau, \bar \tau,\beta) =   \sum_{m=1}^\infty \sum_{\Lambda}{1 \over m}  Z^x_{(1,0)}(\Lambda\tau, \bar \Lambda \tau,m R_x) ~, \label{foldsum}\ee
to recover $Z^x_{cft}(\tau, \bar \tau,R)$ in terms of $Z^x_{(1,0)}(\tau, \bar \tau,R_x)$.

Let us now take a closer look at $Z^x_{(1,0)}(\tau, \bar \tau,\beta)$ which can be written in the form
\be Z^x_{(1,0)}(\tau, \bar \tau,\beta)  = {\beta \over 2 \pi} \int dE_x \, | \eta(\tau)|^{-2} \exp\left[-{2 \pi \tau_2 }  E_x^2 - i  \beta E_x \right] \ . \label{Z10spec1} \ee
If we let
\be E_x = -i (1 - i \epsilon)  E_{\ttau}~, \ee
then one can write
\be Z^x_{(1,0)}(\tau, \bar \tau,\beta)  = -{i \beta \over 2 \pi} \int dE_{\ttau} \, | \eta(\tau)|^{-2} \exp\left[{2 \pi \tau_2 }  E_\ttau^2 -  \beta E_\ttau \right] ~,  \label{Z10spec1a} \ee
With this analytic continuation of the integration variable $E_x$ (which does not change the value of the integral), we see that the path integral expression for the partition function takes on the standard Boltzmann form. This analytic continuation is also interpretable as the path integral of
a timelike non-compact boson, $\ttau=ix$ with action
\be 2 \pi {\cal L} = -\partial \ttau \bar \partial \ttau, \ee
and with  primary operators of the form
\be e^{i E_x x} = e^{-i E_\ttau \ttau},  \ee
whose dimensions and charges are respectively given by
\be \Delta_0 + \bar \Delta_0 = -  E_\ttau^2~,\qquad \Delta_0 - \bar \Delta_0 = 0~,  \qquad \mathfrak{j}_0 + \bar \mathfrak{j}_0 =- 2  i E_\ttau~, \qquad \mathfrak{j}_0 - \bar \mathfrak{j}_0 = 0 \ , \ee
and its descendants.

Now, if one were to imagine first deforming the compact $x$ model and then identifying the $(1,0)$ partition function, one finds
\be Z^{def\ x}_{(1,0)}(\tau, \bar \tau,\beta)  =  -{\beta \over 2 \pi} i\int dE_\ttau \, | \eta(\tau)|^{-2} \exp\left[ {2 \pi \tau_2 \over  l_2^2}  E_\ttau^2 -\beta E_\ttau \right] \ . \label{Z10spec2}\ee

We will now show that (\ref{Z10spec1a}) and (\ref{Z10spec2}) are related by the kernel relation (\ref{c1kernel2}).  To show this, let us first re-write (\ref{Z10spec1a}) in the form
\beq {1 \over \beta} Z^x_{(1,0)}(\tau, \bar \tau, \xi, \bar \xi)& =&  -{1 \over 2 \pi} |\eta(\tau)|^{-2} i \int dE_\ttau \,  e^{-2 \pi \tau_2 (\Delta_0 + \bar \Delta_0)+ 2 \pi i  (\xi \mathfrak{j}_0 - \bar \xi \bar \mathfrak{j}_0)}  \cr
& =&  -{1 \over 2 \pi} |\eta(\tau)|^{-2} i \int dE_\ttau \,  e^{-2 \pi \tau_2 (\Delta_0 + \bar \Delta_0)+  \pi i ( \xi- \bar \xi)(\mathfrak{j}_0 + \bar \mathfrak{j}_0)} \ , \label{Zx10undef} \eeq
where we are relating
\be \label{xibeta}
2 \pi \xi =  {1 \over 2} (\beta \mu + i \beta)~, \qquad 2 \pi \bar \xi = {1 \over 2} (\beta \mu  - i \beta) \ . \ee
Acting on this expression with the kernel  (\ref{c1kernel2}) and changing variables back from $\xi$ to $\beta$ leads to
\be {1 \over \beta} Z^{def\ x}_{(1,0)}(\tau, \bar \tau)  =- i {1 \over 2 \pi} \int dE_\ttau \, | \eta(\tau)|^{-2} \exp\left[ {2 \pi \tau_2 \over  l_2^2}  E_\ttau^2 \right] \ . \label{Z10spec3}\ee
While the modified spectrum $\Delta = \Delta_0/l_2$ is reproduced faithfully, this expression is not useful to apply (\ref{foldsum}) to recover the partition function of the compact $x$ theory, as the dependence on chemical potential $\beta$ is integrated out in (\ref{Z10spec3}). This, however, can be remedied by using the modified kernel discussed in Appendix B of \cite{Hashimoto:2019wct}.  We simply modify (\ref{c1kernel2}) to
$$\fbox{\parbox{\hsize}{\be {1 \over \beta} Z^{def\ x}_{(1,0)}(\tau, \bar \tau, \xi, \bar \xi) = \int {d \xi_0\, d \bar \xi_0 \over \tau_2} {l_2 \over l_1^2} e^{ \pi ((\xi - \xi_0) - (\xi - \bar \xi_0)^2/ \tau_2} e^{- 2 \pi (\xi- \xi_0)(\bar \xi -  \bar \xi_0) / \tau_2 l_1^2 }  {1 \over \beta} Z^x_{(1,0)} (\tau, \bar \tau, \xi_0, \bar \xi_0)~,
\label{c1kernel3}\ee}} $$
and then apply it on (\ref{Z10spec1a}) to obtain $Z^{def\ x}_{(1,0)}(\tau, \bar \tau, \xi, \bar \xi)$. \footnote{Although the kernel formula (\ref{c1kernel3}) has been derived for the case of $J_x\bar J_x$ deformation of a free compact boson, it's likely to be true for $J\bar{J}$ deformation of a larger class of CFT$_2$ with $U(1)_L\times U(1)_R$ global symmetry. In section \ref{sec:sl2r} and \ref{sec:embed}, we apply this kernel formula (\ref{c1kernel3}) to $SL(2,R)$ WZW CFT and consider its string theory embedding. This can be considered as a highly non-trivial check for the validity of the kernel formula beyond free compact boson CFT.}

We can summarize the logic of this subsection using the illustration in figure \ref{figb}. The essential point here is that in order to understand the deformation of the {\it compact theory}, all that is needed is the deformation of the {\it decompactified theory} where the periodicity of the $x$ coordinate is replaced by the {\it chemical potential} in the decompactified theory.  The chemical potential $\beta$ is coupled naturally to the charge associated with the isometry shifting the coordinate $x$ before, and after, the deformation.

\begin{figure}[h]
\centerline{\fbox{\parbox{3.5in}{
$$
\begin{array}{ccc}
 Z_{(1,0)}^{x} (\tau, \bar \tau, \xi, \bar \xi)& \rightarrow  &  Z_{(1,0)}^{def\ x} (\tau, \bar \tau, \xi, \bar \xi)\\
\uparrow & & \downarrow \\
Z_{compact}^{x}(\tau,\bar \tau, \xi,\bar \xi) & {\color{red}\Rightarrow} & Z_{compact}^{def\ x}(\tau,\bar \tau, \xi,\bar \xi)
\end{array}
$$}}}
\caption{Chain of connections relating $Z_{compact}^{x}(\tau,\bar \tau, \xi,\bar \xi)$ and  $Z_{compact}^{def\ x}(\tau,\bar \tau, \xi,\bar \xi)$.  The red arrow is the effect of the deformation that we are primarily interested in understanding. Here, for the compact scalar, we argue that we can obtain the effects of the deformation also using the folding/unfolding track and the application of the kernel. \label{figb}}
\end{figure}

We can systematize this logic even further as follows. We start with a model where $x$ is compact under $x \rightarrow x + \beta$, and the set of primary operators are labeled by integers $(n,w)$ with dimensions and charges
\beq
\Delta_0 + \bar \Delta_0 &= &{4 \pi^2  } {n^2 \over \beta^2} + {1 \over 16 \pi^2}w^2 \beta^2~, \cr
\Delta_0 - \bar \Delta_0 &= &n w ~, \cr
\mathfrak{j}_0 + \bar \mathfrak{j}_0 & =&   {4 \pi  n \over \beta}~,\cr
 \mathfrak{j}_0 - \bar \mathfrak{j}_0 & = & \frac{2\pi w}{\beta}~.
\eeq
Restricting to the $(m,w) = (1,0)$ was seen to be tantamount to decompactifying the $x$ coordinate, which in turn is equivalent to parameterizing
\be n = {\beta E_x \over 2 \pi} ~ , \ee
and approximating the sum over $n$ by an integral over $E_x$ with the measure $(\beta/2\pi) dE_x $, and setting $w=0$. The dimensions and charges in this parameterization becomes
\beq \Delta_0 + \bar \Delta_0 &=&  {E_x^2} \ , \cr
\Delta_0 - \bar \Delta_0 &=&   0 \ , \cr
\mathfrak{j}_0 + \bar \mathfrak{j}_0 & = &2   E_x \ ,\cr
\mathfrak{j}_0 - \bar \mathfrak{j}_0 & = &0 \ .\label{eqnaaa}
\eeq

When we deform the model and consider the gauged action (\ref{sbasicaction2}) with the auxiliary fields, the operators of the deformed theory can be written as
\be e^{i E_x x} e^{-i E_t t}~ ,  \label{dressedx} \ee
and its descendants. The gauge condition (\ref{sgauge}) constrains
\be l_1 E_x+ l_2 E_t = 0 ~. \ee
The dimension and charge of operators (\ref{dressedx}) is a gauge invariant notion. It is convenient to compute it in the $\bar{A}=B=0$ gauge where it is clear that the dimension
\be \Delta + \bar \Delta =  E_x^2 - E_t^2   = {1 \over l_2^2} E_x^2\ . \ee
The charge $E_x$ of the operator (\ref{dressedx}) with respect to the isometry $d/dx$ is independent of the deformation. The partition function $Z^{def\ x}_{(1,0)}(\tau,\bar{\tau})$ which we wish to insert in (\ref{foldsum}) therefore is of the form
\be {1 \over \beta} Z^x_{(1,0)}(\tau, \bar \tau, \xi, \bar \xi)
=  -{1 \over 2 \pi} |\eta(\tau)|^
{-2} i \int dE_\ttau \,  e^{-2 \pi \tau_2 (\Delta + \bar \Delta)+  \pi i ( \xi- \bar \xi)(\mathfrak{j}_0 + \bar \mathfrak{j}_0)} \ , \ee
and is recognizable as the result of applying (\ref{c1kernel3}) to (\ref{Zx10undef}).

\subsection{Embedding into string theory and integrating over $(\tau, \bar \tau)$}

In this subsection, we describe how to embed the worldsheet torus partition function into critical string theory and interpret the result from the spacetime perspective.

Let us begin by writing (\ref{Zx10undef}) in the form

\beq \lefteqn{ {1 \over \beta} Z^x_{(1,0)}(\tau, \bar \tau, \xi, \bar \xi)} \label{Zx10spec}  \\
&=&-{1 \over 2 \pi}  i  \sum_{N, \bar N} d_x(N, \bar N) \int dE_\ttau \,  e^{-2 \pi \tau_2 (\Delta_0 + \bar \Delta_0+N + \bar N -1/12) + 2 \pi i \tau_1(N - \bar N)+  \pi i ( \xi- \bar \xi)(\mathfrak{j}_0 + \bar \mathfrak{j}_0)} \ , \nonumber \eeq
where
\be \sum_{N, \bar N} d_x(N, \bar N) q^N \bar q^{\bar N}\equiv |q^{-1/24}\eta(\tau)|^{-2} ~, \qquad  q\equiv e^{2\pi i \tau}~.\ee

To obtain the one-loop amplitude of the critical string theory, we begin by multiplying the partition functions
\be Z^{\cal N}(\tau, \bar \tau) = \sum_{h, \bar h} d_h(h, \bar h) q^{h-c_{{\cal N}}/24} \bar q^{\bar h-c_{{\cal N}}/24}~, \qquad \tau_2 ^{-1}Z^{ghost}(\tau, \bar \tau) = |\eta(\tau)|^4 \ . \ee
where $\tau_2 ^{-1}Z^{ghost}(\tau, \bar \tau)$ is the quantity (7.2.26) and (7.2.27) in the notation of \cite{Polchinski:1998rq},
so that
\beq
\lefteqn{{1 \over \beta \tau_2}  Z^x_{(1,0)}(\tau, \bar \tau, \xi, \bar \xi) Z^{\cal N}(\tau, \bar \tau)Z^{ghost}(\tau, \bar \tau) } \label{xNghost10} \\
&=&
-{1 \over 2 \pi}  i     \sum_{N, \bar N, h, \bar h} D(N, \bar N, h, \bar h) \int_{-\infty}^\infty dE_\ttau \,  e^{-2 \pi \tau_2 (\Delta_0 + \bar \Delta_0+N+h +\bar N +\bar h - 2)+ 2 \pi i \tau_1 (h - \bar h+N - \bar N) + \pi i ( \xi- \bar \xi)(\mathfrak{j}_0 + \bar \mathfrak{j}_0)} ~,\nonumber \eeq
with
\be \sum_{N, \bar N,h, \bar h} D(N, \bar N, h, \bar h) q^{N+h} \bar q^{\bar N+ \bar h} \equiv (\sum_{N, \bar N} d_x(N, \bar N) q^N \bar q^{\bar N})\times (\sum_{h, \bar h} d_h(h, \bar h) q^{h} \bar q^{\bar h}) \times |q^{-1/24}\eta(\tau)|^4 ~. \ee
The CFT  on ${\cal N}$ is some CFT with $c_{\cal N}=25$ so that the system combined with the CFT $x$ is critical. The sum of $(h, \bar h)$ over the operators of ${\cal N}$ includes the descendants.

To get the one-loop string amplitude, we need to apply (\ref{foldsum}) and sum over $m$ and $\Lambda$, and then integrate $(\tau, \bar \tau)$ over the fundamental domain ${\cal F}$. We can however exchange the sum over $\Lambda$ by integrating $(\tau, \bar \tau)$ over the strip ${\cal S}$ instead. The integral over $(\tau, \bar \tau)$ is interpretable as the level match condition and the Schwinger parameter, and the sum over $m$ is interpretable as the sum over the $m$-string sector of the multi-string gas \cite{Polchinski:1985zf}.

Let us review the technical aspect of the argument of \cite{Polchinski:1985zf} as follows. We start with (\ref{xNghost10}) and first integrate out $E_\ttau$.   This leads to
\beq
\lefteqn{{1 \over \beta \tau_2} Z^x_{(1,0)}(\tau, \bar \tau, \xi, \bar \xi) Z^{\cal N}(\tau, \bar \tau)Z^{ghost}(\tau, \bar \tau) } \label{xNghost10a} \\
&=&
{1 \over 2 \pi}    \sum_{N, \bar N, h, \bar h} D(N, \bar N, h, \bar h)  \sqrt{{1\over 2 \tau_2^2}} e^{-{1 \over 8 \pi  \tau_2}\beta^2  -2 \pi \tau_2 (N+h +\bar N +\bar h - 2)+ 2 \pi i \tau_1 (h - \bar h+N - \bar N)}~. \nonumber \eeq
The integral over $(\tau_1,\tau_2)$ can be done in closed form and leads to
\beq
\lefteqn{{1 \over \beta} \int_{\cal S} {d^2 \tau \over \tau_2} Z^x_{(1,0)}(\tau, \bar \tau, \xi, \bar \xi) Z^{\cal N}(\tau, \bar \tau)Z^{ghost}(\tau, \bar \tau) } \label{xNghost10b} \\
&=&
{1 \over \beta} \sum_{N, \bar N, h, \bar h} D(N, \bar N, h, \bar h)    \left. e^{-\beta \sqrt{(N + \bar N + h + \bar h - 2)}}\right|_{h - \bar h+N - \bar N=0} \ . \nonumber \eeq
This expression is interpretable as the Boltzman sum of single-string on-shell states satisfying the Virasoro and level-matching conditions which will become clear in the discussion that follows. This is what \cite{Polchinski:1985zf} showed in general for string theory of the form $S_1 \times {\cal N}$.

It is useful to understand how one obtains the same conclusion if one were to do the $(\tau, \bar \tau)$ integral first and the $E_{\ttau}$ integral last. To do this, we go back once again to (\ref{xNghost10}) and re-write it as follows
\beq
\lefteqn{{1 \over \beta\tau_2}  Z^x_{(1,0)}(\tau, \bar \tau, \xi, \bar \xi) Z^{\cal N}(\tau, \bar \tau)Z^{ghost}(\tau, \bar \tau) } \label{xNghost10c} \\
&=& {1 \over 2 \pi}  \int_{-\infty}^\infty dy  \int_{-\infty}^\infty  dz \, e^{i z (y - E_{\ttau})} \cr
&&
-{1 \over 2 \pi}  i     \sum_{N, \bar N, h, \bar h} D(N, \bar N, h, \bar h) \int_{-\infty}^\infty dE_{\ttau} \,  e^{-2 \pi \tau_2 (\Delta_0(y) + \bar \Delta_0(y)+N+h +\bar N +\bar h - 2)+ 2 \pi i \tau_1 (h - \bar h+N - \bar N) - \beta E_\ttau}. \nonumber \eeq
The $z$ integral simply inserts a delta function $2 \pi \delta(y-E_{\ttau})$. One can perform the $y$ integral, followed by the $(\tau, \bar \tau)$ integral in closed form, leading to
\beq
\lefteqn{{1 \over \beta}  \int{d^2 \tau \over \tau_2^2} Z^x_{(1,0)}(\tau, \bar \tau, \xi, \bar \xi) Z^{\cal N}(\tau, \bar \tau)Z^{ghost}(\tau, \bar \tau) } \label{xNghost10d} \\
&=& \sum_{N, \bar N, h, \bar h} D(N, \bar N, h, \bar h)  \int_{-\infty}^\infty dE_{\ttau} \, \left.  e^{-\beta E_{\ttau}}\int_{-\infty}^\infty dz\,  {e^{-i E_{\ttau} z + i z  \sqrt{N+h + \bar N + \bar h-2}} \over z} \right|_{h - \bar h+N - \bar N=0}. \nonumber \eeq
The $z$ integral leads to a step function $\Theta(E_{\ttau} -  \sqrt{N+h + \bar N + \bar h-2})$. Finally, by integrating by parts, we arrive at
\beq
\lefteqn{{1 \over \beta}  \int{d^2 \tau \over \tau_2^2} Z^x_{(1,0)}(\tau, \bar \tau, \xi, \bar \xi) Z^{\cal N}(\tau, \bar \tau)Z^{ghost}(\tau, \bar \tau) } \label{xNghost10d} \\
&=& {1 \over \beta} \sum_{N, \bar N, h, \bar h} D(N, \bar N, h, \bar h)  \int_{-\infty}^\infty dE_{\ttau} \, \left.  e^{-\beta E_{\ttau}} \delta(E_{\ttau} -  \sqrt{N+h + \bar N + \bar h-2}) \right|_{h - \bar h+N - \bar N=0}. \nonumber \eeq
It's easy to see that the boundary term arising from the integration by part vanishes.
The analysis above manifests that the $(\tau, \bar \tau)$ integral imposes the Virasoro and level matching condition which constrains the zero mode integral over $E_{\ttau}$ to its on-shell value.  We will take advantage of this perspective as follows. The effect of $J_x \bar J_x$ deformation simply modifies the spectrum $(\Delta, \bar \Delta)$ of each state keeping $(\mathfrak{j}_0, \bar \mathfrak{j}_0)$ unchanged.  The density of off-shell states parameterized by $E_{\ttau}$ are unaffected by the deformation. The effect of the deformation on the single string one-loop zero point amplitude can be understood as stemming from the change in the Virasoro condition arising strictly from the change in the dimensions $(\Delta, \bar \Delta)$ of the operators corresponding to the single string states. Taking advantage of this perspective, one can make manifest that the one-loop zero point amplitude with compact Euclidean time is the Boltzmann sum of the on-shell string states in the $SL(2,R)$ case as we will show below.

Summing over multi-string sectors labeled by $m$ in (\ref{foldsum}) will recover the full multi-string one-loop zero point amplitude. (The sum over $\Lambda$ is not needed since we integrated $(\tau, \bar \tau)$ over ${\cal S}$.)

In the discussion that follows, we will implement the kernel formula on the $SL(2,R)$ WZW model and eventually embed it into critical string theory. This, as we will show, has the interpretation of  $J_3\bar{J_3}$ deformation of string theory in global-$AdS_3\times \mathcal{N}$ which from the spacetime point of view can be viewed as ``single-trace $T\bar{T}$ deformation" of the boundary  CFT$_2$.

\section{$J_3 \bar J_3$ deformation of $SL(2,R)$ and $H_3$ sigma model \label{sec:sl2r}}

In this section, we will extend the analysis of the $J_3 \bar J_3$ deformation to $SL(2,R)$ and $H_3$ sigma model. The target space of $SL(2,R)$ sigma model is $AdS_3$, and $H_3$ is its continuation. The goal of this exercise is to derive the partition function of the sigma model on worldsheet torus with $H_3$ as the target space with compact Eucleidan time coordinate. By embedding the $H_3$ into critical string theory on $H_3 \times {\cal N}$, we can interpret the result as the thermal partition function of the spacetime theory. Unlike the free boson warm-up exercise of the previous section, the effects of $J_3 \bar J_3$ deformation are much more intricate. However, we can follow the template of the analysis of the free boson and obtain a concrete expression for the partition function of the deformed $H_3$ sigma model.

The spectrum of operators on $SL(2,R)$ sigma model and the worldsheet torus partition function on $H_3$ was analyzed in \cite{Maldacena:2000hw,Maldacena:2000kv}. The spacetime interpretation by embedding $SL(2,R)$ and $H_3$ into critical string theory, $AdS_3 \times {\cal N}$ and $H_3 \times {\cal N}$, have also been analyzed in \cite{Maldacena:2000hw,Maldacena:2000kv}. The goal of this section is to generalize the result of these analyses to the $J_3 \bar J_3$ deformed model.

We will begin by reviewing the basic background of $SL(2,R)$  and $H_3$ sigma models mostly reviewing \cite{Maldacena:2000hw,Maldacena:2000kv}. After reviewing the known facts about the $SL(2,R)$ and $H_3$, we will provide a precise formulation of the $J_3 \bar J_3$ deformed models and compute data such as its spectrum and the torus partition function. Just like in the compact scalar case, the deformed partition function can be expressed in terms of an integral kernel form. We will eventually embed the deformed sigma model into critical string theory and comment on the physical features of the deformed spacetime theory.

\subsection{Review of $SL(2,R)$ sigma model}

In this subsection, we review the physics of the $SL(2,R)$ WZW model.  We will mostly follow \cite{Maldacena:2000hw,Maldacena:2000kv,Maldacena:2001km}.  As the subject itself is rather technical, our review will mostly consist of establishing conventions with selected comments.

Group manifold $SL(2,R)$ is a Lorentzian signature geometry. There are various ways to parameterize $g \in SL(2,R)$. One example is
\be g = e^{{i \over 2} \theta_L \sigma_3} e^{{1 \over 2} \rho \sigma_1} e^{{i \over 2} \theta_R \sigma_3}~, \ee
and if we further parameterize
\be \label{thetalr}
\theta_L =\ttau+ \phi~, \qquad \theta_R =\ttau- \phi ~,\ee
the metric takes the form
\be ds^2 =k \left( - \cosh^2 (\rho/2) d \ttau^2 + {1 \over 4} d \rho^2 + \sinh^2 (\rho/2)  d \phi^2 \right) ~,\label{ads3metric} \ee
and describes a Lorentzian-filled cylinder geometry with $AdS_3$ metric. The existence of isometries $d/d \ttau$ and $d/d \phi$ are manifest in this expression. There will be a corresponding conserved quantum number that we will identify shortly. There is also a $B$-field and a constant dilaton.

The WZW action of a Lie group manifold takes the form
\be S =   -{k \over 2 \pi} \int d z_0\,  dz_1\ \mbox{Tr} ( \partial g^{-1} \bar \partial g )+ k \Gamma_{WZ} ~,\ee
where $k$ denotes the level of the Lie algebra and also encodes the Neveu-Schwarz three-form field strength $H$,
in terms of which the action takes the form
\beq S &=& {k \over 4 \pi}\int dz_0 \, dz_1 \, \left( \partial \rho \bar \partial \rho - \partial \theta_L \bar \partial \theta_L - \partial \theta_R \bar \partial \theta_R -  \cosh\rho (\partial \theta_R \bar \partial \theta_L+\partial \theta_L \bar \partial \theta_R)\right)  \cr
&& - {k \over 4 \pi}\int dz_0 \, dz_1 \, (B_0 + \cosh\rho-1) (\partial \theta_R \bar \partial \theta_L-\partial \theta_L \bar \partial \theta_R) \label{SL2Rsigma}~, \eeq
where $B_0$ is the additive constant component of the Neveu-Schwarz 2-form potential in the $\tau \phi$ directions which does not affect the $H=dB$ field.\footnote{Except at $\rho=0$ where there is a delta function localized $H$ flux.} We will be setting $B_0=0$  for most of this paper, \footnote{This is also required form maintining $SL(2,R) \times SL(2,R)$ invariance of the global $AdS_3$ background.} and we will have more to say about the meaning of the parameter $B_0$ in section \ref{secB0}.

We will consider the conserved holomorphic, anti-holomorphic currents\footnote{See (A.20) of \cite{Bufalini:2021ndn} where the relation between ${\bf J}_3$ and $J_3$ are explained.}
\beq {\bf J}_3(z) &=&{k \over 2} \mbox{Tr} \left( i \sigma_3 \partial g g^{-1}\right) =  -{k \over 2} ( \partial \theta_L + \cosh \rho  \partial \theta_R) \equiv - i J_3(z)~,  \cr
\bar {\bf J}_3(\bar z)  &=& {k \over 2}  \mbox{Tr} \left( i \sigma_3 g^{-1} \bar \partial g \right)  =- {k \over 2} (\bar \partial \theta_R +\cosh \rho \bar \partial \theta_L) \equiv - i \bar J_3(\bar z)~, \eeq
which are normalized such that
\be J_3(z) J_3(0) \sim - {k \over 2}{1 \over z^2} \ , \label{Jnorm} \ee
and as such they are timelike.

The spectrum of operators of this $SL(2,R)$ sigma model was analyzed in \cite{Maldacena:2000hw}. The operators
\be V^{j;w}_{m,\bar m} = \Phi^j_{m, \bar m} e^{\sqrt{{2 \over k}}[(m+{k \over 2} w) Y_L + (\bar m + {k \over 2} w) Y_R]} ~,\label{Vjmmw} \ee
are labeled by the quantum numbers $j$, $m$, $\bar m$, and the spectral flow parameter $w$. $\Phi^j_{m,\bar m}$ are not charged with respect to $J_3$ and $\bar J_3$, and
\be J_3 = - \sqrt{{k \over 2}} \partial Y_L~, \qquad \bar J_3 = - \sqrt{{k \over 2}} \bar \partial Y_R \ . \ee

The dimensions and charges of operators are defined by the standard relation
\be T(z)  V^{j;w}_{m,\bar m}(0) = {\Delta_0 \over z^2} V^{j;w}_{m,\bar m}(0)+ \ldots\ , \qquad \bar T(\bar z)  V^{j;w}_{m,\bar m}(0) ={\bar \Delta_0 \over \bar z^2} V^{j;w}_{m,\bar m}(0) + \ldots\ , \ee
\be J_3(z)  V^{j;w}_{m,\bar m}(0) = {\mathfrak{j}_0 \over z}V^{j;w}_{m,\bar m}(0) + \ldots, \qquad \bar J_3(\bar z)  V^{j;w}_{m,\bar m}(0) ={\bar \mathfrak{j}_0 \over \bar z} V^{j;w}_{m,\bar m}(0) + \ldots\ .
\ee
Here, the subscript ``$0$'' is to indicate the fact that these are the spectrum data for the undeformed $SL(2,R)$ sigma model. The vertex operators (\ref{Vjmmw}) have dimensions and charges \cite{Maldacena:2000hw}
\beq \Delta_0 &=& -{j(j-1) \over (k-2)} + {m^2 \over k} - {1 \over k }\left(m + {k w \over 2}\right)^2 ~,  \cr
\bar \Delta_0 &=& -{j(j-1) \over (k-2)} + {\bar m^2 \over k} - {1 \over k }\left(\bar m + {k w \over 2}\right)^2  ~, \cr
\mathfrak{j}_0 &=& \left(m + {k w \over 2} \right)~, \cr
\bar \mathfrak{j}_0 &=& \left(\bar m + {k w \over 2} \right) \ . \label{undefDelK}
\eeq
The charges
\be E\equiv \mathfrak{j}_0 + \bar \mathfrak{j}_0= m + \bar m + k w~, \qquad l\equiv \mathfrak{j}_0 - \bar{\mathfrak{j}}_0=m - \bar m  \in Z ~,\label{j0charges} \ee
are interpretable as the charges associated with the isometries $d/d \ttau$ and $d/d \phi$ respectively.

One key feature of the $SL(2,R)$ sigma model explained in \cite{Maldacena:2000hw} is that the operators (\ref{Vjmmw}) come in two categories. One is referred to as ``Discrete States'' and are parameterized by $(j, n, \bar n, w,  N, \bar N)$ where \footnote{Here we focus on the $\mathcal{D}^+_j$ sector.}
\be m = j + n~, \qquad \bar m = j + \bar n~, \ee
with
\be {1 \over 2} \leq j \leq {k - 1 \over 2}~, \ \text{ and } \  n, \bar n = \mbox{non-negative integers}~.  \ee
$w$ is an integer-valued spectral flow parameter, and $N,\bar{N}$ are the left and right moving $SL(2,R)$ oscillator numbers. There are some degeneracy $d(n, \bar n, w, N, \bar N)$ for fixed $(n, \bar n, w, N , \bar N)$.  The parameter $j$ is continuous at the level of considering the set of $SL(2,R)$ operators.  (This parameter is discretized when $SL(2,R)$ sigma model is embedded into critical string theory through the Virasoro constraint.)

The other, referred to as ``Continuum States,'' on the other hand, are parameterized by $(s, \alpha, n, \bar n, w, N, \bar{N})$,
with
\be j = {1 \over 2} + i s~, \qquad m = \alpha + n~, \qquad \bar m = \alpha + \bar n ~. \ee
The parameter $\alpha$ takes value in range $0 \le \alpha < 1$. One can think of ${\alpha, n, \bar n}$ being parametrized by one continuous parameter $0 < \alpha + n< \infty$ and one discrete parameter $l = n - \bar n$. The existence of continuum states can be understood as arising from the existence of long strings \cite{Maldacena:1998uz,Seiberg:1999xz}. Note that for the continuous states, $j$, $m$, and $\bar{m}$ are not related, unlike the discrete states.

To embed the $SL(2,R)$ sigma model into critical string theory, we consider the $SL(2,R) \times {\cal N} \times \mbox{(ghost)}$ where the central charges  $c_{SL(2,R)}+c_{\cal N}$ of the matter, add up to 26 in the case of bosonic string theory. The spectrum of free strings from the spacetime point of view is parameterized by the $SL(2,R)$ operator data and the dimension $(h, \bar h)$ of the CFT  ${\cal N}$, subjected to the Virasoro constraints:
\be \Delta_0  +N + h = 1 ~, \qquad \bar \Delta_0 + \bar  N + \bar h = 1 \ . \ee

Finally, the one-loop partition function was computed in \cite{Maldacena:2000kv} using an earlier work \cite{Gawedzki:1991yu}.
What \cite{Maldacena:2000kv} considered is the worldsheet torus partition function for the sigma model (\ref{SL2Rsigma}) analytically continued to $H_3 = SL(2,C)/SU(2)$ such that
\be\ttau = i x \label{tix}~, \ee
so that the non-linear sigma model metric is
\be ds^2 =  k \left(\cosh^2 (\rho/2) dx^2 + {1 \over 4} d \rho^2 + \sinh^2 (\rho/2)  d \phi^2 \right)~, \label{H3metric}\ee
with periodicities \footnote{This background is that of thermal $AdS_3$ and/or Euclidean BTZ.}
\be (x, \phi) \sim (x , \phi + 2 \pi) \sim (x + \beta, \phi + \beta \mu)\ .  \label{periodicity} \ee
The resulting partition function, therefore, is a function of the worldsheet moduli $(\tau, \bar \tau)$ and the spacetime moduli data $\beta$ and $\mu$. The result of the analysis, given in eq.(27) of  \cite{Maldacena:2000kv}, is
\be Z^{H_3}_{compact}(\beta, \mu; \tau, \bar \tau) = \sum_{n,m} Z^{H_3}_{(m,n)}(\beta, \mu; \tau, \bar \tau)={\beta (k-2)^{1/2} \over 2 \pi \sqrt{\tau_2}} \sum_{(m,n)} {e^{-k \beta^2 |m-n \tau|^2/4 \pi \tau_2 + 2 \pi (Im\,  U_{n,m})^2/\tau_2} \over |\theta_1(\tau,U_{n,m})|^2} \label{ZH_3} ~,\ee
with
\be U_{m,n} = -{i \over 2 \pi} (\beta + i \mu \beta) (n \tau - m)~, \qquad \bar U_{m,n} = {i \over 2 \pi} (\beta - i \mu \beta) (n \bar \tau - m) \ . \ee

It was further shown in eq.(61)-(63) of \cite{Maldacena:2000kv}, by embedding the $Z^{H_3}_{compact}(\beta, \mu; \tau, \bar \tau) $ partition function in critical string theory and performing an intricate $(\tau,\bar \tau)$ integral,  that the single string contribution to the one-loop string amplitude\footnote{Here, $(h, \bar h)$ parameterizes the set of all operators in internal space, ${\cal N}$, including the descendants.}
\beq  \Xi^{H_3}_{(1,0)}(\beta,\mu) & = &   \int_{{\cal S}} {d \tau_1 d \tau_2 \over \tau_2^2} Z_{(1,0)}^{H_3} (\beta, \mu; \tau, \bar \tau) Z_{{\cal N}} (\tau, \bar \tau) Z_{ghost} (\tau, \bar \tau) \label{thermalpart} \\
& =&  \sum_{h, \bar h, N, \bar N, w}  D(h, \bar h, N, \bar N, w) \left[ \sum_{n, \bar n} e^{-\beta E - i \beta \mu l} + \int_0^\infty ds\, \rho(s) e^{-\beta E(s) - i \beta \mu l} \right]~, \nonumber \eeq
with
\be \rho(s) = -{1 \over  \pi} \log(\epsilon) + {1 \over 4 \pi i} {d \over ds} \log \left( {\Gamma({1 \over 2} - i s + \bar m) \Gamma({1 \over 2} - i s  - m) \over \Gamma({1 \over 2} + i s  + \bar m) \Gamma({1 \over 2} + i s  - m)} \right)~,  \label{rhos}\ee
and
\be m = - {k \over 4} w + {1 \over w}\left({s^2 + {1 \over 4} \over k-2} + N + h - 1 \right), \qquad
\bar m = - {k \over 4} w + {1 \over w}\left({s^2 + {1 \over 4} \over k-2} + \bar N + \bar h - 1 \right)~, \label{mms}\ee
is interpretable as the spacetime thermal partition function of single string states analogous to \cite{Polchinski:1985zf}. The analysis of \cite{Maldacena:2000kv} can be understood as the generalization of  \cite{Polchinski:1985zf} where the compact Euclidean time coordinate $x$  is warped non-trivially along the radial direction as can be seen in (\ref{H3metric}).

As a consequence of the presence of the continuum states, the conformal field theory on $H_3$ is non-compact even when the Euclidean time coordinate is compactified. This is what gives rise to the $\log \epsilon$ divergence \cite{Maldacena:2000kv} in (\ref{rhos}).

This concludes the review of the analysis of $SL(2,R)$ and $H_3$ sigma models by \cite{Maldacena:2000hw,Maldacena:2000kv}. Next, we will describe its $J_3 \bar J_3$ deformation.

\subsection{$J_3 \bar J_3$ deformation of the $SL(2,R)$ sigma model \label{sec:J3J3def}}

In this subsection, we will formulate $J_3 \bar J_3$ deformed $SL(2,R)$ sigma model as a null gauged sigma model $SL(2,R) \times R_t \times U(1)_y / (U(1)_L \times U(1)_R)$ of the type discussed \eg\ in \cite{Bars:1991pt,Quella:2002fk,Israel:2003ry,Israel:2004ir,Martinec:2020gkv,Bufalini:2021ndn}.\footnote{It is equivalent to a timelike coset $SL(2,R) \times R_t / U(1)$, similar to the equivalence of (\ref{aaa}) and (\ref{bbb}).}

The gauged action takes the form\footnote{As was the case in the previous section,  regard $A_\alpha,B_\alpha$ as independent gauge fields even though $A,\bar{B}$ do not appear in (\ref{basicaction}).}
\beq \lefteqn{ 2 \pi {\cal L}} \cr
 & =&   {k \over 2} \left( \rule{0ex}{3ex}\partial  \rho \bar \partial \rho - \partial \theta_L (\bar \partial \theta_L + 2 l_1 \bar A) - (\partial \theta_R + 2 l_1 B) \bar \partial \theta_R - 2 (\partial \theta_R + l_1 B) (\bar \partial \theta_L + l_1 \bar A) \cosh \rho \right. \cr
 && \qquad \left. \rule{0ex}{3ex} + (\partial \theta_R \bar \partial \theta_L-\partial \theta_L \bar \partial \theta_R) \right) \cr
&& -  ( \partial t \bar \partial t - 2l_2 B\bar \partial t -2 l_2 \bar A  \partial t + 2 l_2^2 B \bar A ) \cr
&& +  \partial y \bar \partial y - 2 l_3 B\bar \partial y + 2 l_3 \bar A  \partial y - 2 l_3^2 B \bar A \ .   \label{basicaction}
\eeq
We will see below that such an action defines a one-parameter family of conformal field theories, and we will
  take (\ref{basicaction}) as defining the $ J_3 \bar J_3$ deformation parameterized by the deformation parameter $l_1$.

First, we note that this action is invariant  under gauge transformation (up to total derivative) \cite{Israel:2004ir}
\beq \label{fulltrans2}
\theta_L(z, \bar z) &\rightarrow & \theta_L(z, \bar z) +l_1 \Lambda(z, \bar z) ~,\cr
\theta_R(z, \bar z) &\rightarrow & \theta_R(z, \bar z) + l_1 M(z, \bar z)~, \cr
t(z, \bar z) & \rightarrow& t(z, \bar z) - l_2 \Lambda(z, \bar z) - l_2 M(z, \bar z) ~,\cr
y(z, \bar z) & \rightarrow & y(z, \bar z) + l_3 \Lambda(z, \bar z) - l_3 M(z, \bar z)~, \cr
A(z,\bar z) & \rightarrow & A(z,\bar z) - \partial \Lambda(z,\bar z)~, \\
\bar A(z, \bar z) & \rightarrow & \bar A(z, \bar z) - \bar \partial \Lambda(z, \bar z) ~,\cr
B(z, \bar z) & \rightarrow &  B(z,\bar z) - \partial M(z, \bar z) ~,\cr
\bar B(z, \bar z) & \rightarrow & \bar B(z, \bar z) - \bar \partial M(z, \bar z) \nonumber ~,\eeq
with \footnote{We stress that $B_0$ which appeared in (\ref{SL2Rsigma}) is set to zero here. The term proportional to $B_0$ is not invariant under (\ref{fulltrans2}). This is related to a footnote on page 4 of \cite{Giveon:1991jj}. \label{fn11}}
\be -{k \over 2} l_1^2 -l_2^2 +l_3^2 = 0 \ .  \ee
We will frequently find it convenient to parameterize
\be \label{lnullcond}
l_2 = \sqrt{1 - {k l_1^2 \over 2}}~, \qquad l_3= 1 ~  . \ee
The gauge fields couple to currents
\be J = l_1 J_3 + l_2 J_t +  l_3 J_y~, \qquad  \bar J = l_1 \bar J_3 + l_2 \bar J_t - l_3 \bar J_y \ . \label{currents} \ee
We take $l_1 < 0$, $l_2 > 0$, and $l_3 >0$ so that the coordinates $(\ttau,\phi)$ and $(t,y)$ have the same orientation when projected to the $J = \bar J = 0$ sector.

As in section \ref{sec:scalar}, the variable $y$ is compact with periodicity $y\sim y+2\pi R_y$.

One feature easily derivable from (\ref{basicaction}) is the geometry of the deformed sigma model target space. This is achieved by integrating out $\bar A$ and $B$ while using (\ref{fulltrans2}) to set some of the embedding coordinates to zero as is done \eg\ in \cite{Dijkgraaf:1991ba,Israel:2004ir,Martinec:2018nco,Martinec:2017ztd}.
This is a version of the temporal/static gauge for which the Faddeev-Popov determinant is trivial.  One convenient choice is to set $\theta_L = \theta_R = 0$.  This leads to a sigma model whose target space geometry is\footnote{There is also a non-trivial dilaton in the background, unlike the $SL(2,R)$ case.}
\beq \label{tausigzerogauge}
ds^2 &=& - {1 \over 2} \left(1 - {2 l_2^2 \over \Delta} \right) d t^2 + {1 \over 2} \left(1 - {2l_3^2 \over \Delta} \right) d y^2 + {k \over 4}  d \rho^2  ~,\cr
B & = & \left( -{l_2 \over l_3}  + {2 l_2 l_3  \over \Delta} \right) dt \wedge dy ~,\\
e^{2 \Phi-2\Phi_0} & = & {2 \over \Delta} ~,\nonumber
  \eeq
where
\begin{equation}
    \Delta = {k  l_1^2 \over 2} \cosh \rho + l_2^2 + l_3^2~,
\end{equation}
and $\Phi_0$ is the value of the dilaton field at $\rho=0$.
The other obvious choice of gauge fixing condition is set $t = y = 0$. This leads to a sigma model whose target space geometry is
\beq \label{ty0gauge}
ds^2 & = & -{k \cosh^2(\rho/2)  \over 1+ \varepsilon \cosh^2(\rho/2) } d \ttau^2 + {k(1 + \varepsilon) \sinh^2(\rho/2) \over 1 + \varepsilon \cosh^2(\rho/2)} d \phi^2 + {k \over 4} d \rho^2 ~,\cr
B & = & {k \sinh^2(\rho/2) \over 1 + \varepsilon \cosh^2(\rho/2)} d \ttau \wedge d \phi ~, \\
e^{2 \Phi- 2 \Phi_0} & = & \frac{1+\varepsilon}{1 + \varepsilon \cosh^2 (\rho/2)}~, \nonumber \eeq
where \footnote{The  $ \varepsilon<0$ deformation is obtained by  a vector gauging in the $J_3,\bar{J}_3$ direction instead of the axial one in (\ref{basicaction}),(\ref{currents}).}
\be \varepsilon = {k l_1^2 \over 2 l_2^2} \ ,  \label{varep}  \ee
and the coordinates $\ttau,\phi$ defined in (\ref{thetalr}).
The Euclidean continuation of the non-linear sigma model constructed using (\ref{ty0gauge}) is the $H_3$ counterpart of (\ref{VV0ratio}) encountered previously for the free compact boson.

We have set  $B=0$  at $\rho=0$ in (\ref{tausigzerogauge})  and (\ref{ty0gauge}) by shifting the constant part which does not affect $H = d B$.  \footnote{This is also required by gauge invariance (see footnote \ref{fn11}).}

These two choices of gauge from the sigma model point of view are related from the target space point of view simply as coordinate transformation
\be t =-{2 l_2 \over l_1}\ttau, \qquad y = -{2 \over l_1} \phi~.
 \label{coordinatemap}\ee
The relative minus sign is a consequence of the fact that we work in the branch where $l_1 < 0$, $l_2 > 0$, and $l_3=1$ in (\ref{currents}).

Another feature that is manifest in (\ref{tausigzerogauge}) and (\ref{ty0gauge}) is the fact that there are two isometries. In  (\ref{tausigzerogauge}), they are $d/dt$ and $d/dy$. In  (\ref{ty0gauge}), they are $d/d \ttau$ and $d/d\phi$. Clearly, the isometries in two gauge choices are related by coordinate transformation (\ref{coordinatemap}). These isometries imply that the operators/states in the CFT can be labeled by the eigenvalues associated with these isometries.

Regardless of the choice of gauge/coordinates, the target space geometry approaches  $AdS_3$ in the small $\rho$ limit whereas in the large $\rho$ limit, the geometry approaches a linear dilaton geometry $R_t \times R_\rho \times S^1$ where $\rho$ is the linear dilaton coordinate. The undeformed limit $l_1 \rightarrow 0$ is such that the large $\rho$ asymptotics becomes $AdS_3$. In the maximally deformed $l_2 \rightarrow 0$ limit, the target space geometry becomes that of Lorentzian ``time'' times Euclidean cigar, $R_t \times (SL(2,R) /U(1))$. The $t=y=0$ gauge is more convenient for taking the limit $l_1 \rightarrow 0$ to recover the undeformed $SL(2,R)$ sigma model as a limit. The $\theta_L=\theta_R=0$ gauge is more convenient for taking the $l_2 \rightarrow 0$ limit where the sigma model approaches the  $R_t \times  (SL(2,R) /U(1))$  limit.

We can also see that because of (\ref{coordinatemap}) and the fact that $\phi$ is $2 \pi$ periodic,
$y$ must have periodicity $y \sim y + 2 \pi R_y$, where
\be R_y = -{2 \over l_1} \ .  \label{Ryconst} \ee
One can easily understand this periodicity of $y$ as the consequence of the fact that large gauge transformations of the form
\be \Lambda(z, \bar z) =- {1 \over l_1} z_1~, \qquad M(z, \bar z) = {1 \over l_1} z_1~, \label{largegauge} \ee
where  $z_1$ is defined in (\ref{z1z2}), which  shifts $\phi \rightarrow \phi+2 \pi$, also shifts $y \rightarrow y - 4 \pi l_1^{-1}$. For (\ref{largegauge}) to be a consistent large gauge transformation, the periodicity of the $y$ coordinate is constrained to (\ref{Ryconst}).

Another feature that can easily be extracted from the $t=y=0$ gauge fixed sigma model with $\bar A$ and $B$ integrated out is that in the small $l_1$ expansion with (\ref{lnullcond}) imposed, the sigma model  (\ref{ty0gauge}) has the form
\be 2 \pi {\cal L}_{def} = 2 \pi {\cal L}_{SL(2,R)} - l_1^2  J_3 \bar{J}_3 + \ldots \label{def123} \ee
and as such it establishes near the $l_1=0$ undeformed limit that this exactly marginal deformation is generated by the $J_3 \bar{J}_3$ deformation.\footnote{Indeed,  (\ref{ty0gauge}) was obtained by $J_3\bar{J}_3$ deformation of $SL(2,R)$ \cite{Giveon:1993ph}.} This is precisely analogous to the observation that the action shifted by $l_1^2 J_x \bar J_x$ in the scalar case (\ref{VV0ratio}). \footnote{As before, $J_3\bar{J}_3$ deformation in the opposite direction of (\ref{def123}) is obtained by a vector instead of an axial gauging in the $J_3,\bar{J}_3$ direction.}

\subsection{Statement of problem \label{sec:problem}}

At this point, we can provide a mathematically precise formulation of the problem of computing the torus partition function of the $J_3 \bar J_3$ deformed $H_3$  that we will be pursuing in this paper.
\begin{enumerate}
\item We consider the CFT action in the form of the gauged sigma model (\ref{basicaction}) and analytically continue
\be \ttau = i x \ . \ee
\item Impose periodicity to $x,\phi$ fields according to (\ref{periodicity}).
\item Compute the worldsheet torus partition function of $J_3\bar{J}_3$ deformed $H_3$, $Z_{compact}^{def\ H_3} (\beta, \mu; \tau, \bar \tau, l_1)$. This quantity will depend on the worldsheet moduli $(\tau, \bar \tau)$, the spacetime moduli $\beta$ and $\mu$, and the deformation parameter $l_1$.
\end{enumerate}
The problem is identical to considering a non-linear sigma model whose target space geometry is the Euclidean continuation of (\ref{ty0gauge}) with periodicity (\ref{periodicity}). The $l_1=0$ limit will reduce the problem to $H_3$ which has already been solved \cite{Maldacena:2000kv}.

The problem is mathematically well-defined but appears difficult on the first pass. In the remainder of this section, we will provide a solution in the form of an integral expression which we will obtain using the fact that (1) the deformed action has a concrete gauged sigma model form,  (2) the folding/unfolding trick of \cite{Polchinski:1985zf} can be utilized, and the fact that the $l_1=0$ limit is already solved in \cite{Maldacena:2000kv}. The analysis will closely follow the template established for the case of a free compact boson in the previous section.

\subsection{Spectrum of operators in $J_3 \bar J_3$ deformed $SL(2,R)$ theory}

In this subsection, we enumerate the operators of the gauged sigma model (\ref{basicaction}) where the target space has a Lorentzian signature with a non-compact time coordinate. To do this, it is convenient to work in the $\bar A = B = 0$ gauge. This gauge will cause the $SL(2,R)$ sector, the $t$, and the $y$ sector in (\ref{basicaction}) to decouple. One can therefore construct the vertex operators as if they are the tensor product of $SL(2,R)$, $t$, and $y$ sectors. The only effect of the gauging stems from the residual gauge symmetries associated with the gauge parameters $\Lambda(z, \bar z)$ and $M(z,\bar z)$ being constants. This will constrain the set of zero mode quantum numbers of the vertex operators to equivalence classes.

The primary vertex operator takes the form
\be   V^{j;w}_{m,\bar m} e^{-i E_t t + i P_L y_L + i P_R y_R}~, \label{Vdressed} \ee
where $V^{j;w}_{m,\bar m}$ (\ref{Vjmmw}) is the vertex operator in the $SL(2,R)$ sector,
\be P_{L,R} = {n_y \over R_y} \pm {w_y R_y \over 2}~, \ee
are the momentum and winding quantum numbers on compact $y$ coordinate, and $E$ is the continuous momentum quantum number on the non-compact $t$ coordinate.

With the proper condition (\ref{Ryconst}) on the periodicity of the $y$ coordinate, the large gauge transformation (\ref{largegauge}) shifts $(w, w_y) \rightarrow (w+1, w_y+1)$ keeping $w-w_y$ fixed. We can use this gauge freedom to set $w_y=0$.

In order for the vertex operators (\ref{Vdressed}) to make sense in the gauged model, it must be invariant under the residual gauge transformation by constant $\Lambda$ and $M$. This then imposes a constraint
\beq 0 & = &  l_1 \left(m + {k \over 2} w \right) + l_2 E_t + {n_y \over R_y}~,  \cr
0 & = &   l_1 \left(\bar m + {k \over 2} w \right) + l_2 E_t -{n_y \over R_y}  ~, \label{chargeconstraint} \eeq
on the zero mode quantum numbers. It is easy to see that the above constraints (\ref{chargeconstraint}) can be obtained by demanding that the vertex operators (\ref{Vdressed}) are annihilated by the null gauge currents (\ref{currents}) with (\ref{lnullcond}).  The Faddeev-Popov determinant from the $\bar A = B = 0$ partial gauge choice will cancel two towers of oscillator modes coming from the $t$ and $y$ sectors. So we only need to consider the descendent of $ V^{j;w}_{m,\bar m}$ in  (\ref{Vdressed}) for the full theory.

The constraints  (\ref{chargeconstraint}) implies that we can solve for  $(E_t,n_y)$ and parameterize the operators by $(m, \bar m)$, or solve for $(m, \bar m)$ and parameterize the operators by $(E_t, n_y)$. This choice is another manifestation of gauge choice which leads to target space geometries (\ref{tausigzerogauge}) and (\ref{ty0gauge}).  Clearly, the two descriptions are physically equivalent, and it would be useful to be able to go back and forth between the two descriptions. This is facilitated by recalling that, equations (\ref{coordinatemap}) and  (\ref{chargeconstraint}) construct the dictionary.

It is now straightforward to compute the dimension of the operator (\ref{Vdressed}). Simply think of it as the $SL(2,R)$ operator parameterized by $(j, w, m, \bar m)$ that is dressed by plane wave operator $e^{-i E_t t + P_L y_l + P_R y_R}$. So its dimensions are
\beq \Delta &=& \Delta_0  - {E_t^2 \over 2} + {P_L^2 \over 2} ~,\cr
\bar \Delta &=& \bar{\Delta}_0  - {E_t^2 \over 2} + {P_R^2 \over 2}~. \eeq
Recalling that the charges are constrained by (\ref{chargeconstraint}), we arrive at
\beq \Delta &=&\Delta_0 + {l_1^2 \over 8} (\mathfrak{j}_0 - \bar \mathfrak{j}_0)^2 - {l_1^2 \over 8 l_2^2} (\mathfrak{j}_0 + \bar \mathfrak{j}_0)^2 ~, \cr
\bar \Delta &=&\bar \Delta_0 + {l_1^2 \over 8} (\mathfrak{j}_0 - \bar \mathfrak{j}_0)^2 - {l_1^2 \over 8 l_2^2} (\mathfrak{j}_0 + \bar \mathfrak{j}_0)^2 \ ,  \label{defdel1}
\eeq
where $(\Delta_0,\bar{\Delta}_0,\mathfrak{j}_0,\bar{\mathfrak{j}}_0)$ are given in (\ref{undefDelK}).

Finally, let us address the global symmetry and the associated charges of the model (\ref{basicaction}). We can show that the deformed model (\ref{basicaction}) admits holomorphic and anti-holomorphic currents ${\cal J}$ and $\bar {\cal J}$ which we construct as follows.

Let's define
\be {\cal J} = r_1 J_3 + r_2 J_t + r_3 J_y ~, \qquad \bar {\cal J} = r_1 \bar J_3 + r_2 \bar J_t - r_3 \bar J_y~. \ee
In the gauge where we set $w_y=0$, the ${\cal J}$ and $\bar{\cal J}$ charges respectively read
\be \mathfrak{j}  = r_1 \left( m + {k \over 2}w\right) + r_2 E_t + r_3 {n_y \over R_y} ~ , \qquad \bar \mathfrak{j} = r_1 \left(\bar m + {k \over 2}w\right) + r_2 E_t - r_3 {n_y \over R_y} ~. \label{deformedJ} \ee
Using (\ref{chargeconstraint}) to eliminate $E_t$ and $n_y$ in terms of $m$ and $\bar m$, plugging it into (\ref{deformedJ}) and imposing the orthogonality condition
\be J(z){\cal J}(0)\sim 0  \ \Rightarrow  \ {k \over 2} l_1 r_1 + l_2 r_2 - l_3 r_3 = 0~, \ee
and a normalization convention
\be {\cal J}(z) {\cal J}(0) = -{{k \over 2} r_1^2 + r_2^2 - r_3^2 \over z^2}= -{k \over 2}{1 \over z^2}~, \label{calJnorm} \ee
leads to
\be \mathfrak{j}+ \bar \mathfrak{j} = {1 \over l_2}(\mathfrak{j}_0 + \bar \mathfrak{j}_0)~ , \qquad
\mathfrak{j} - \bar \mathfrak{j} = l_2(\mathfrak{j}_0 - \bar \mathfrak{j}_0) \ . \ee
Here, (\ref{calJnorm}) is one arbitrary convention for establishing the normalization of ${\cal J}$ and $\bar {\cal J}$.

What we learn from this exercise is that while ${\cal J}$ and $\bar {\cal J}$ are holomorphic currents with eigenvalues $\mathfrak{j}$ and $\bar \mathfrak{j}$, their linear combinations
\be \mathfrak{j}_0 + \bar \mathfrak{j}_0 = l_2 (\mathfrak{j} + \bar \mathfrak{j})~, \qquad \mathfrak{j}_0 - \bar \mathfrak{j}_0 = {1 \over l_2} (\mathfrak{j} - \bar \mathfrak{j})~,  \label{j0j} \ee
are the charges with respect to the manifest isometries $d/d \ttau$ and $d/d\phi$. In other words, by combining the holomorphic and anti-holomorphic currents ${\cal J}$ and $\bar {\cal J}$, we can construct the Noether current associated with $d/d \ttau$ and $d/d\phi$ for the deformed theory.\footnote{Unlike the $(\mathfrak{j},\bar \mathfrak{j})$, $(\mathfrak{j}_0,\bar \mathfrak{j}_0)$ are not charges of holomorphic and anti-holomorphic currents of the deformed theory. Nonetheless, $\mathfrak{j}_0 + \bar\mathfrak{j}_0 = l_2 (\mathfrak{j}+\bar\mathfrak{j})$ constitutes the charge of a non-holomorphic Noether current  $d / d \ttau$.} This confirms the expectation that vertex operators (\ref{Vdressed}) subject to (\ref{chargeconstraint}) are also equipped with quantum numbers associated with these isometries since these isometries were manifestly preserved in the non-linear sigma model data (\ref{tausigzerogauge}) and (\ref{ty0gauge}).

The fact that the CFT defined by (\ref{basicaction}) admits holomorphic and anti-holomorphic currents ${\cal J}$ and $\bar {\cal J}$ is a strong statement that may prove useful for extracting more information about this CFT.
To address the problem stated in section \ref{sec:problem}, the relation (\ref{j0j})  relating charges associated with the isometry generators, $d/d \ttau$ and $d/d\phi$, and the holomorphic-antiholomorphic generators, ${\cal J}$ and $\bar {\cal J}$, is the main information that we need to extract.

 We close this subsection by summarizing, below, the dimensions and charges of the vertex operators of the gauge sigma model (\ref{basicaction}):
\be \fbox{\parbox{4in}{\beq \Delta &=&\Delta_0 + {l_1^2 \over 8} (\mathfrak{j}_0 - \bar \mathfrak{j}_0)^2 - {l_1^2 \over 8 l_2^2} (\mathfrak{j}_0 + \bar \mathfrak{j}_0)^2 ~,\cr
\bar \Delta &=&\bar \Delta_0 + {l_1^2 \over 8} (\mathfrak{j}_0 - \bar \mathfrak{j}_0)^2 - {l_1^2 \over 8 l_2^2} (\mathfrak{j}_0 + \bar \mathfrak{j}_0)^2~, \cr
 \mathfrak{j}_0+ \bar \mathfrak{j}_0 &=&  l_2(\mathfrak{j} + \bar \mathfrak{j}) ~,\cr
 \mathfrak{j}_0- \bar \mathfrak{j}_0& =& {1 \over l_2}(\mathfrak{j} - \bar \mathfrak{j})~, \nonumber
\eeq}} \label{spec}
\ee
where  $(\Delta_0,\bar{\Delta}_0,\mathfrak{j}_0,\bar{\mathfrak{j}}_0)$ are given in (\ref{undefDelK}). We will use this data (\ref{spec}) to construct the one-loop thermal partition function of the gauge sigma model (\ref{basicaction}) in the subsequent subsections.

\subsection{$J_3 \bar J_3$ deformation as a kernel \label{seckern}}

In this subsection, we formulate the partition function of the $J_3 \bar J_3$ deformed CFT as an integral kernel of the undeformed CFT for the $SL(2,R)$ model generalizing the previous statement (\ref{c1kernel2})  for the $c=1$ model.

Consider a  Boltzmann factor for a state in the undeformed $SL(2,R)$ model
\be \exp\left[2 \pi i (\tau (\Delta_0+ N - c/24) - \bar \tau (\bar \Delta_0+ \bar N - c/24) )+ 2 \pi i \xi \mathfrak{j}_0 - 2 \pi i \bar \xi \bar \mathfrak{j}_0 \right]  \ . \label{SL2RBoltzman} \ee
This is a Boltzmann factor in the worldsheet sense, where $(\tau, \bar \tau)$ encodes the temperature and the twist of the worldsheet theory, and $(\xi, \bar \xi)$ are the chemical potential for charges $\mathfrak{j}_0$ and $\bar \mathfrak{j}_0$.
The quantities $(\Delta_0, \bar \Delta_0, \mathfrak{j}_0, \bar \mathfrak{j}_0)$ are as is given in (\ref{undefDelK}).
The full $SL(2,R)$ partition function is a sum of these Boltzmann factors in a spectral decomposition. We will elaborate on the structure of this sum later, but for the time being, let us consider the contribution of a single state.

One would like to formally derive the expression analogous to (\ref{c1kernel2}) for the $SL(2,R)$ model, but here, we will take the approach of making an educated guess and justifying it post priory.\footnote{We present a separate argument motivating the form of the kernel for (\ref{SL2RBoltzman}) in Appendix A.}

Consider an integral transform of (\ref{SL2RBoltzman}) of the form
$$ \fbox{\parbox{\hsize}{
\beq&&  \int {d \xi\, d \bar \xi \over \tau_2}  {l_2 \over l_1^2}e^{- \pi k  (\xi - \bar \xi)^2/2\tau_2} e^{{- \pi  \xi \bar \xi / \tau_2 l_1^2}} \cr
&& \qquad   \ \exp\left[2 \pi i (\tau (\Delta_0+ N - c/24) - \bar \tau (\bar \Delta_0+ \bar N - c/24) ) + 2 \pi i \xi \mathfrak{j}_0 - 2 \pi i \bar \xi \bar \mathfrak{j}_0 \right]\cr
& \equiv & \exp\left[2 \pi i (\tau (\Delta+ N - c/24) - \bar \tau (\bar \Delta+ \bar N - c/24) )\right] ~.  \label{SL2RBoltzKern}
\eeq}}$$
The kernel in the first line of (\ref{SL2RBoltzKern}) is identical to (\ref{c1kernel2}) if we set $k=-1$. One can think of the form of the kernel as being motivated by the universality of $J \bar J$ deformation but for general $k$ similar to what was discussed in appendix A of  \cite{Hashimoto:2019wct}. The last line of (\ref{SL2RBoltzKern}) is the result of doing the $(\xi, \bar \xi)$ integral which is Gaussian and can be done in closed form.

The fact that the result of the $(\xi, \bar \xi)$ integral given in the last line of (\ref{SL2RBoltzKern}) is also in the Boltzmann form is quite non-trivial by itself. The values of $\Delta$ and $\bar \Delta$ in the last line (\ref{SL2RBoltzKern}) are constrained by the first two lines, but it turns out to be precisely the expression (\ref{defdel1}) which we found in the analysis of the spectrum of deformed $SL(2,R)$ model.

This establishes that the kernel (\ref{SL2RBoltzKern}) will deform the Boltzmann factor in the spectral decomposition of a partition function term by term. This is a rather powerful statement which we will be utilizing extensively to address the problem formulated in section \ref{sec:problem}.

\subsection{Folding and unfolding procedure for the $H_3$ model \label{secfold}}

In the previous subsection, we obtained an explicit expression for an integral transform (\ref{SL2RBoltzKern}) which converts the Boltzmann factor of the undeformed $SL(2,R)$ model to that of the $J_3 \bar J_3$ deformed model. On the other hand, the problem we posed in section \ref{sec:problem} is to derive the $J_3 \bar J_3$ deformation of the Euclidean compactified $H_3$ model. Fortunately, we can follow the same folding/unfolding procedure used in section \ref{sec:fold}. {\it{The main idea is that the partition function on compactified Euclidean $H_3$ is related to the partition function of Lorentzian decompactified $SL(2,R)$ with dependence on chemical potential along the lines of (\ref{Z10spec1}). }}

One important difference between the compact scalar in section \ref{sec:fold} and $H_3$ is that in $H_3$, there are two isometries $d/dx$ and $d/d \phi$. The $\phi$ coordinate is $2 \pi$ periodic. The folding/unfolding trick only involves the $x$ coordinate, but there is one more data that parameterizes the twist in the $\phi$ coordinate when we compactify the $x$ coordinate.

To see this issue more clearly, consider a partition function of two free bosons whose target space is a torus with  complex structure moduli (in ($x, \phi)$ coordinates)
\be ( \mu \beta + i \beta), \
\text{and } \  ( \mu \beta - i \beta) ~ .\label{xibeta2} \ee
The zero mode contribution to the partition function (excluding the Casimir term $-c/12$), in this case, as given in (2.4.11) of \cite{Giveon:1994fu}, takes the following form \footnote{Equation (\ref{HHH}) is in the convention that $\alpha'=2$ whereas (2.4.1) of \cite{Giveon:1994fu} was in the convention that $\alpha'=1$. As such, $g = 2G$ and $b = 2B$.}
\be\label{HHH}
 H = \left[ n_i (g^{-1})^{ij} n_j + {1 \over 4} m^i(g - b g^{-1} b)_{ij} m^j +  m^i b_{ik} (g^{-1})^{kj} n_j \right]~, \ee
with
\be g = \left(\begin{array}{cc}1 &{\beta \mu \over 2 \pi} \\ { \beta \mu \over 2 \pi}& {\beta^2+\beta^2 \mu^2 \over 4 \pi^2} \end{array}\right)~, \qquad b=i \left(\begin{array}{cc}0 &  {\beta b_0 \over \pi} \\ -  { \beta b_0 \over \pi} & 0 \end{array}\right)~, \ee
and $i,j=1,2$.
Here, $(n_2, m_2)$ are the winding numbers of the worldsheet torus mapping onto the  Euclideanized time coordinate. If one Poisson resumes $n_2$ to $w_2$ and then set $(w_2,m_2) = (1,0)$, and then integrate in $E_{\ttau}$ as was done in (\ref{Z10spec1a}), we obtain
\be 2 \pi \tau_2 H = 2 \pi \tau_2\left( n_1^2 + {1 \over 4} m_1^2\right) + i \beta \mu n_1 - 2 \pi \tau_2  E_{\ttau}^2 + \beta (E_{\ttau}+ b_0 m_1) \ . \ee
Although this analysis was for the case where target space $T^2$ is a decoupled sector of the worldsheet theory, this property should continue to hold if the $T^2$ is warped as in the case for $AdS_3$ (\ref{ads3metric}) and its deformation (\ref{ty0gauge}).

What we learn from this exercise, in analogy to (\ref{Zx10spec}), is the statement that
\beq \lefteqn{{1 \over \beta} Z_{(1,0)}^{H_3}(\tau, \bar \tau, \xi, \bar \xi) }\cr & = &
 \sum_{states} e^{-2 \pi \tau_2 (\Delta_0 + \bar \Delta_0+ N + \bar N -c/12) + 2 \pi i \tau_1 (\Delta_ 0- \bar \Delta_0 + N - \bar N) - \beta E_{\ttau}
- i \mu \beta l} \cr
&=&
 \sum_{states} e^{-2 \pi \tau_2 (\Delta_0 + \bar \Delta_0+ N + \bar N -c/12)+ 2 \pi i \tau_1 (\Delta_ 0- \bar \Delta_0 + N - \bar N) + \pi i (\xi - \bar \xi)(\mathfrak{j}_0 + \bar \mathfrak{j}_0)
- \pi i (\xi + \bar \xi)(\mathfrak{j}_0 - \bar \mathfrak{j}_0)} ~,\label{Z10H3}
\eeq
and
\beq \lefteqn{{1 \over \beta} Z_{(1,0)}^{def\ H_3}(\tau, \bar \tau, \xi, \bar \xi)}  \cr & =  &
 \sum_{states} e^{-2 \pi \tau_2 (\Delta + \bar \Delta+ N + \bar N -c/12) + 2 \pi i \tau_1 (\Delta- \bar \Delta + N - \bar N)- \beta E_{\ttau}
- i \mu \beta l} \cr
&=&
 \sum_{states} e^{-2 \pi \tau_2 (\Delta + \bar \Delta+ N + \bar N -c/12) + 2 \pi i \tau_1 (\Delta- \bar \Delta + N - \bar N) + \pi i (\xi - \bar \xi)(\mathfrak{j}_0 + \bar \mathfrak{j}_0)
- \pi i (\xi + \bar \xi)(\mathfrak{j}_0 - \bar \mathfrak{j}_0)} ~,\label{Z10H3def}
\eeq
with
\be 2 \pi \xi =  \beta \mu + i \beta~, \qquad 2 \pi \bar \xi = \beta \mu  - i \beta \ . \label{xibetamu} \ee
We are also using the fact that
\be \mathfrak{j}_0 + \bar \mathfrak{j}_0 \equiv E_{\ttau}~, \qquad \mathfrak{j}_0 - \bar \mathfrak{j}_0 = m - \bar m \equiv l \in Z~, \label{eeetau}\ee
are the charges (\ref{j0charges}) associated with $d/d \ttau$ and $d/d\phi$. The normalization of (\ref{xibetamu}) differs by a factor of 2 compared to (\ref{xibeta}) because $\mathfrak{j}_0 + \bar \mathfrak{j}_0$ is normalized differently betweeen (\ref{eeetau}) and (\ref{eqnaaa}). Note that the energy $E_t$ in  (\ref{Vdressed}) is the charge associated with the isometry generator $d/d t$ and is relateded to $E_\ttau$ defined in (\ref{eeetau})  via the dictionary
\be R_y E_t = -R_y { l_1 \over 2 l_2} E_{\ttau} =  {1 \over l_2} E_{\ttau} = {1 \over l_2}(\mathfrak{j}_0 + \bar \mathfrak{j}_0) = \mathfrak{j} + \bar \mathfrak{j}~.\label{EEtaut} \ee
The above relation can be easily read off from the coordinate transformation (\ref{coordinatemap}).

We will be more precise about the meaning of sum over ``states" when we write the detailed form of the spectral decomposition of $Z_{(1,0)}^{H_3} (\tau, \bar \tau, \xi, \bar \xi) $ or $Z_{(1,0)}^{def\ H_3}(\tau, \bar \tau, \xi, \bar \xi)$ in section \ref{sec:embed}, but at this point, a useful point to stress is that they are the same sums for both the deformed and the undeformed theory. The only difference between (\ref{Z10H3}) and (\ref{Z10H3def}) is that one has $\Delta_0$ and $\bar \Delta_0$ whereas the other has $\Delta$ and $\bar \Delta$.

This concludes our quest to collect all the necessary data we may need to answer the question posed in section \ref{sec:problem}, which we do in the following section.

\subsection{Answer to the problem posted in section \ref{sec:problem} \label{sec:answer}}

We are now ready to combine the result of section \ref{seckern} and section \ref{secfold} to formulate the answer to the problem posed in section \ref{sec:problem}.  This can be summarized as follows.

\begin{enumerate}
\item Explicit expression for $Z_{(1,0)}^{H_3}$ can be inferred from (\ref{ZH_3})

\be Z^{H_3}_{(1,0)}(\beta, \mu; \tau, \bar \tau) = {\beta (k-2)^{1/2} \over 2 \pi \sqrt{\tau_2}}   {e^{-(k-2) \beta^2 /4 \pi \tau_2} \over |\theta_1(\tau,-i(\beta-i \mu \beta)/2\pi)|^2} \label{ZH310}~. \ee

\item $Z_{(1,0)}^{def H_3}$ can be inferred from $Z_{(1,0)}^{H_3}$ using the integral transform
\be \fbox{\parbox{5.25in}{\beq \lefteqn{ {1 \over \beta} Z_{(1,0)}^{def\ H_3}(\tau, \bar \tau, \xi, \bar \xi, l_1)} \cr & =& \int {d \xi_0\, d \bar \xi_0 \over \tau_2} {l_2 \over l_1^2} e^{ \pi ((\xi - \xi_0) - (\xi - \bar \xi_0)^2/ \tau_2} e^{-  \pi (\xi- \xi_0)(\bar \xi -  \bar \xi_0) / \tau_2 l_1^2 }  {1 \over \beta} Z_{(1,0)}^{H_3}(\tau, \bar \tau, \xi_0, \bar \xi_0)
\nonumber \ , \eeq}}  \label{H3kernel3}\ee
where  $\beta$, $\mu$, $\xi$, and $\bar \xi$ are related according to (\ref{xibetamu})
\be 2 \pi \xi = \beta \mu + i \beta~, \qquad 2 \pi \bar \xi =  \beta \mu  - i \beta \ . \ee
This is what was discussed in section \ref{seckern}.

\item $Z^{def\ H_3}_{compact}$ can be obtained from $Z_{(1,0)}^{def\ H_3}$ by applying (\ref{foldsum})

\be Z^{def\ H_3}_{compact}(\beta, \mu, \tau, \bar \tau,l_1) =   \sum_{m} \sum_{\Lambda}{1 \over m}  Z^{def\ H_3}_{(1,0)}(m\beta, \mu ; \Lambda\tau, \bar \Lambda \tau, l_1) \ . \label{foldsum2}\ee
This is what was discussed in section \ref{secfold} and then summing over the modular images generated by the action of $\Lambda\in SL(2,Z)/Z$ and $m$.

\end{enumerate}
We can also summarize this procedure by a flow diagram analogous to figure \ref{figb}

\begin{figure}[h]
\centerline{\fbox{\parbox{3.5in}{
$$
\begin{array}{ccc}
 Z_{(1,0)}^{H_3} (\tau, \bar \tau, \xi, \bar \xi)& \rightarrow  &  Z_{(1,0)}^{def\ H_3} (\tau, \bar \tau, \xi, \bar \xi)\\
\uparrow & & \downarrow \\
Z_{compact}^{H_3}(\tau,\bar \tau, \xi,\bar \xi) &{\color{red}\Rightarrow} & Z_{compact}^{def\ H_3}(\tau,\bar \tau, \xi,\bar \xi) \ .
\end{array}
$$}}}
\caption{Chain of connections relating $Z_{compact}^{H_3}(\tau,\bar \tau, \xi,\bar \xi)$ and  $Z_{compact}^{def\ H_3}(\tau,\bar \tau, \xi,\bar \xi)$. This is the $H_3$ version of the flow diagram illustrated in figure \ref{figb} for the compact scalar. The red arrow is the deformation we are interested in computing. In the $H_3$ case, we do not presently have the means to compute the deformation associated with the red arrow directly, but we can follow the black arrow and obtain the result that we are after albeit in a somewhat roundabout manner.
\label{figb2}}
\end{figure}

These three steps above are the answer to the problem posed in section \ref{sec:problem} and the main result we wish to present in this paper.

The partition function of the  $J_3 \bar J_3$ and other related exactly marginal deformation of $SL(2,R)$ WZW model are also presented in equation (5.8) of  \cite{Israel:2003ry} for the case where $\beta=0$.  It is important to compare (\ref{H3kernel3}) with (5.8) of  \cite{Israel:2003ry}. Both expressions involve integral over zero modes but their equivalence is not manifest. We hope to return to this issue in the future, but for now, we will proceed to discuss the consequences of  (\ref{H3kernel3}) in the rest of this paper by embedding the deformed sigma model in critical string theory.

\section{String theory on $(J_3 \bar J_3$ deformed $AdS_3) \times {\cal  N}$ \label{sec:embed}}

In this section, we discuss various physical spacetime features of $J_3 \bar J_3$ deformed $SL(2,R)$ and $H_3$ sigma models after embedding them into critical string theory.  To study the on-shell spacetime dynamics, we will consider the Lorentzian $SL(2,R)$ version of the story. In order to study the spacetime vacuum one-loop amplitude as the thermal partition function, we will consider the $H_3$ story.

In sections \ref{sec:41} to \ref{sec4.4}, we will go over various on-shell physics in the Lorentzian signature frame.

\subsection{Semi-classical analysis of the long strings\label{sec:41}}

One of the important special features of string theory on $AdS_3 \times {\cal N}$ with NSNS flux is the existence of long strings \cite{Maldacena:1998uz,Seiberg:1999xz}. At the semi-classical level (\ie\  $ k\to \infty$), one can understand these objects by considering the embedding of strings winding the periodic $\phi$ coordinate at some fixed $\rho$. We will take the trivial embedding into a point in ${\cal N}$ so that it does not contribute energetically. The existence of the long string state is a consequence of the fact that the effective potential experienced by the string,
\be U_{eff}(\rho) = {1 \over 2 \pi} \int_0^{2 \pi} d \phi \, \left( |w|\sqrt{-g_{\ttau \ttau} g_{\phi \phi}} - w B_{\ttau \phi} \right)\ , \ee
approaches a constant value in the large $\rho$ limit for $w \ge 1$. This feature is unique to $AdS_3$. Large branes can exist for $AdS$ in other dimensions, but it requires turning on angular momentum \cite{Hashimoto:2000zp}.

It is straightforward to do the same analysis for the $J_3 \bar J_3$ deformed target spacetime (\ref{tausigzerogauge}). Since only the $AdS_3$ part of the spacetime is deformed and ${\cal N}$ is factorized, we only need the information contained in (\ref{tausigzerogauge}) or (\ref{ty0gauge}). These two background geometries are related by a coordinate transformation (\ref{coordinatemap}), so they contain the same information. We choose to compute the energy in $t$ coordinate in units of $R_y$ which is natural in (\ref{tausigzerogauge}) coordinate system
\be R_y E_t^{probe} = \left.{R_y \over 2 \pi \alpha'} \int_0^{2 \pi R_y} dy \, \left( |w|\sqrt{-g_{tt} g_{yy}} - w B_{ty} \right) \right|_{\rho=\infty} ={1 \over 2 \lambda} (|w| - l_2 w) ~,\label{EprobeS}\ee
where we recall (\ref{EEtaut})
\be R_y E_t = -R_y { l_1 \over 2 l_2} E_{\ttau} =  {1 \over l_2} E_{\ttau} = {1 \over l_2}(\mathfrak{j}_0 + \bar \mathfrak{j}_0) = \mathfrak{j} + \bar \mathfrak{j}~, \ee
are various equivalent ways of parameterizing the energy, and
\be \lambda \equiv {1 \over 2}l _1^2 = {\alpha' \over  R_y^2}~, \label{a1klam}\ee
is a convenient parameterization of the deformation parameter.

\begin{figure}
\centerline{
\begin{tabular}{cc}
\includegraphics[width=3in]{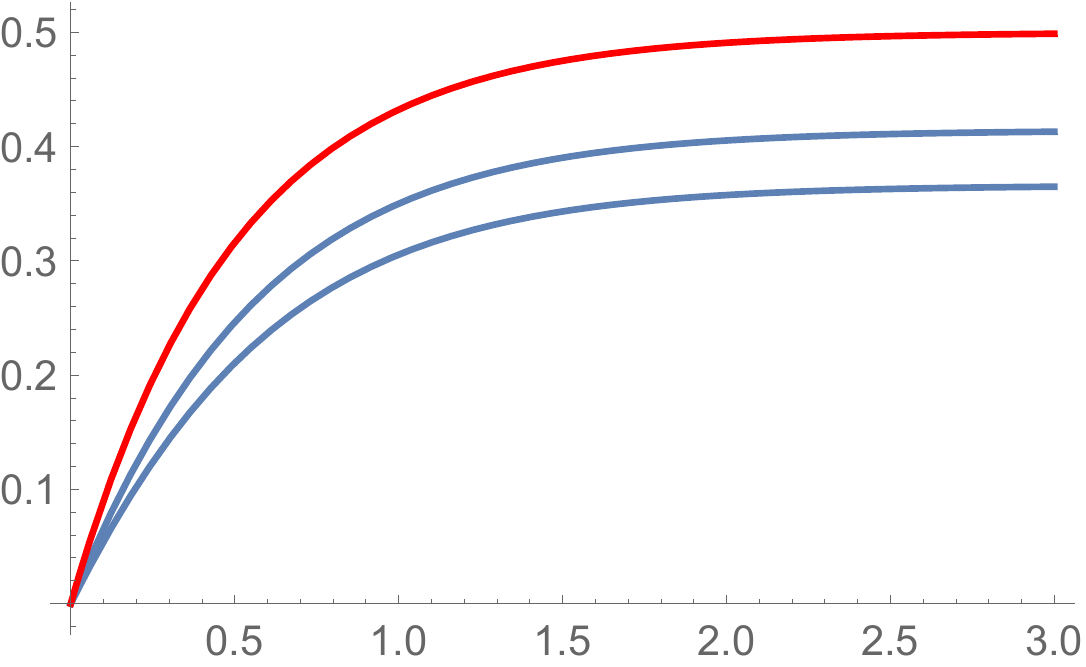}&  \includegraphics[width=3in]{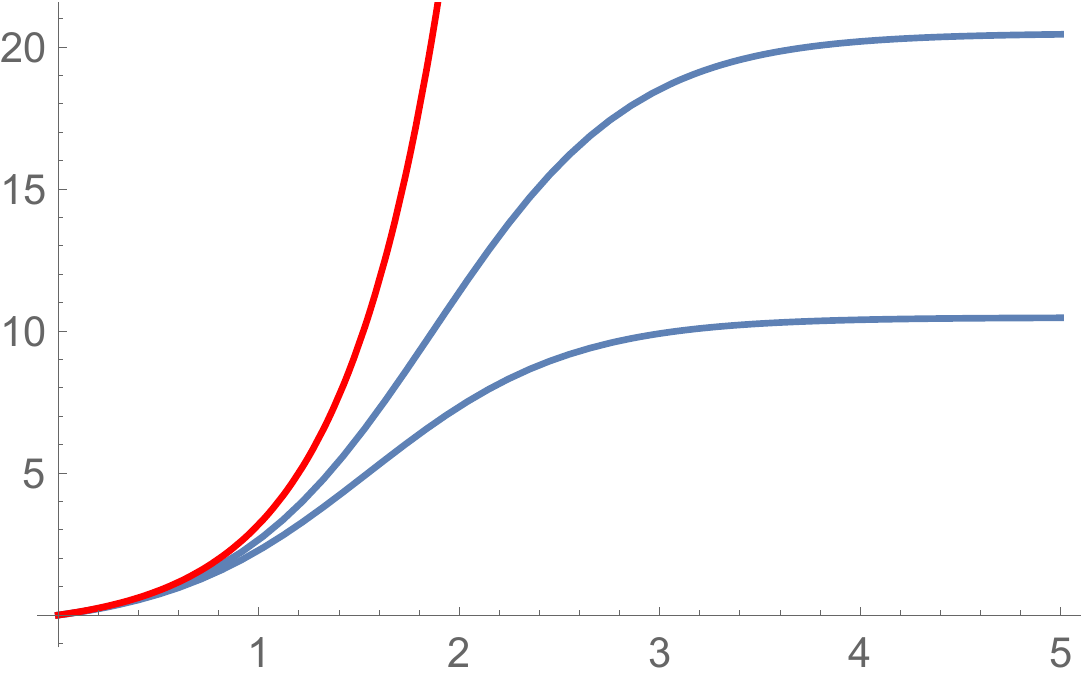} \\
(a) & (b)
\end{tabular}
}
\caption{$R_y E_t^{probe}(\rho)$ as a function of $\rho$ for a long string with winding (a) $w=1$ and (b) $w=-1$. The red line is the undeformed $AdS_3$ limit.  The blue lines are for increasing values of the $J_3 \bar J_3$ deformation parameter $\lambda$. The $w=-1$ states become infinitely massive and decouple in the $AdS_3$ limit, but do have finite energy in the large $\rho$ limit for any finite deformation. \label{figa}}
\end{figure}

In figure \ref{figa}a and \ref{figa}b, we plot $R_y E_t^{probe}(\rho)$ for $w=1$ and $w=-1$, respectively. The red curve corresponds to the undeformed limit. We see that for $w=1$, the long string that exists for $AdS_3$ continues to exist as $J_3 \bar J_3$ deformation is turned on. Perhaps more interesting is the observation that in the $J_3 \bar J_3$ deformed theory, the long string exists also for $w=-1$, but the value of $R_y E_t^{probe}(\rho = \infty)$ goes to infinity in the undeformed limit. This implies that long strings exist with both positive and negative winding in the $J_3 \bar J_3$ deformed theories, but the ones with negative winding become infinitely massive and decouple in the $AdS_3$ limit (see also \cite{Giveon:2017myj}). This is not unexpected because in the cigar limit $l_2 \rightarrow 0$, long strings exist with both positive and negative windings.

\subsection{Comment on constant $B_{ty}$ \label{secB0}}

In this subsection, we will briefly discuss the issue of the possibility of adding a constant  $B_{ty}^0 = B_{ty}(\rho=0)$  in  (\ref{tausigzerogauge}). This modification will not affect $H = dB$ and as such the supergravity equations of motion are not affected.

The constant additive $B^0_{ty}$ will, however, modify (\ref{EprobeS}) by an amount proportional to $w$. In other words, $B^0_{ty}$ will act like a chemical potential to winding number $w$. It is however strange for there to be a chemical potential for a charge $w$, which is not conserved in $AdS_3$. We will in fact argue that the natural value for $B^0_{ty}$ is zero as follows. From the spacetime point of view, $\rho=0$ is not a special point. It is simply the origin of a local polar coordinate system parameterized by $\rho$ and $y$. A non-vanishing $B_{ty}$ would then imply that there is a non-vanishing $H \sim \delta^2(\rho)  dt \wedge dy \wedge d\rho$ NSNS 3-form  field strength. \footnote{Such a singular $H$-field signifies the presence of a localized source in the middle of the geometry which empty $AdS_3$ and its deformation considered in this paper shouldn't have.}  It is not immediately clear if string theory in the presence of this singular $H$ field can be considered sensible.

Recently, a similar issue regarding the constant component of NSNS $B$-field was discussed in \cite{Martinec:2023plo} (see also \cite{Ashok:2021ffx}). The discussion in \cite{Martinec:2023plo} was mostly in the context of BTZ black holes but when continued to the global $AdS_3$ limit, the condition on $B_{ty}$ that \cite{Martinec:2023plo} found to be natural (see between (2.11) and (2.12) in \cite{Martinec:2023plo})  is in agreement with setting $B_{ty}$ to zero. \footnote{It is curious that in the finite temperature case, preferred $B$-field \cite{Martinec:2023plo} appears to lead to a delta function supported $H$ field at the origin in the Euclideanized background. This issue is presumably connected to subtle aspects of the black hole information paradox. This goes beyond the scope of what we are prepared to discuss here.} In the deformed theory, the issue of the constant $B$-field was discussed in some detail in \cite{Chakraborty:2024ugc}.

To avoid introducing a singular three-form $H$-field strength at $\rho=0$, we set  $B_{ty}^0=0$. We will find that the energy spectrum of continuum states in the semiclassical limit is in agreement with (\ref{EprobeS}) partly justifying our choice to set $B_{ty}^0=0$.  \footnote{Recall that this was imposed by gauge invariance in footnote \ref{fn11}.}

\subsection{Spectrum of string states \label{sec:43}}

In this subsection, we will describe the spectrum of closed string states that arises upon $J_3 \bar J_3$ deforming the  $AdS_3 \times {\cal N}$ spacetime. In $AdS_3 \times {\cal N}$, there are states arising from the discrete and continuous representations of $SL(2,R)$ as was described in \cite{Maldacena:2000hw}. These states have their counterparts in the deformed theory. We will describe some of the features that we can infer easily using the results of the earlier sections of this note.

For both the continuum and the discrete states, we will be imposing the Virasoro constraint
\be \Delta  +N+ h = 1~, \qquad \bar \Delta+\bar{N} + \bar h = 1~, \label{viscons} \ee
where $\Delta$ and $\bar \Delta$ is what we worked out in (\ref{spec}) and is explicitly dependent on the deformation parameter $l_1$. $(N,\bar N)$ are the oscillator number in the $SL(2,R)$ sector. The oscillator number in ${\cal N}$ is contained in the internal-space scaling dimensions $(h, \bar h)$.
Here, we are working in bosonic string theory so that $c_{{\cal N}} = 26  - c_{SL(2,R)}$.  Working instead on superstrings will lead to minor modifications in the formulas presented below.  To compute the spectrum of the spacetime theory, one also needs to impose the null gauge constraints (\ref{chargeconstraint}) when necessary.

\subsubsection{Continuum States \label{seccont}}

The tracking of continuum states under the $J_3 \bar J_3$ deformation is a bit easier. We begin by recalling that on $AdS_3\times \mathcal{N}$, continuous states correspond to a state created by a product  vertex operator (\ref{Vjmmw}) and a vertex operator of ${\cal N}$, with
\be j = {1 \over 2} + i s~, \qquad m = \alpha + n~, \qquad \bar m = \alpha + \bar n \ . \ee
As such, the continuum states are labeled by continuous parameters  $s$, $\alpha$, and integer parameters $w$,  $n$,  $\bar n$, $N$, and $\bar N$, possibly with some degeneracy. Upon imposing the Virasoro constraints one can easily compute the spacetime spectrum of the long string states as done in \cite{Maldacena:2000hw}.

Upon $J_3\bar{J}_3$ deformation of $AdS_3\times \mathcal{N}$, the primary vertex operators of the deformed theory are given by (\ref{Vdressed}) subject to the null gauge constraints (\ref{chargeconstraint}) times a vertex operator of  ${\cal N}$ and then imposing the Virasoro constraints (\ref{viscons}) with (\ref{spec}) and (\ref{undefDelK}). To compute the spacetime spectrum, it is often useful to look at the sum and the difference of Virasoro constraints. The difference
\be \Delta - \bar \Delta+N-\bar{N}   + h -\bar  h = 0~, \ee
leads to constraints on discrete data $w$, $n$, $\bar n$, $N$,  $\bar N$, and $h -\bar  h$. This constraint turns out to be insensitive to the $J_3 \bar J_3$ deformation. The sum, on the other hand, leads to the constraint
\be \Delta + \bar \Delta+N+\bar{N} + h +\bar  h  = 2 \ . \label{vsum} \ee
We can use this constraint to fix, for instance, the continuous parameter $\alpha$ in terms of the rest of the data.

We focus on energy
\be  R_y E_t =\mathfrak{j} + \bar \mathfrak{j} \label{EK} \ . \ee
Recall that $l_1$ is negative, so $E_t$ is positive, (\ref{Ryconst}). One can compute the energy of a state satisfying the Virasoro constraint by solving (\ref{vsum}) and substituting it into $j$ as is given in (\ref{spec}),(\ref{undefDelK}). This will lead to a somewhat complicated expression but can be made to look simple by doing the following. First, note that in the $l_1 \rightarrow 0$ limit, the complicated expression reduces to
\be R_y E_t^0 \equiv {1 \over w} \left(- {2 j(j-1) \over k-2} + { k w^2 \over 2} + N + \bar N + h + \bar h - 2 \right) \ . \label{RE0} \ee
This expression is equivalent to (80) of \cite{Maldacena:2000hw}.  Now solving for $R_y E_t$ from (\ref{vsum}) and using (\ref{RE0}) one obtains
\be R_y E_t=
 \sqrt{\frac{w^2}{4\lambda
   ^2}+\frac{w (R_y  E_t^0-kw/2)}{\lambda }+(n-\bar n )^2} - {l_2 w \over 2 \lambda}~. \label{defE} \ee
This is a remarkable formula, and for $w=1$,  it almost looks like the spectrum of a $T \bar T$ deformed CFT \cite{Smirnov:2016lqw,Cavaglia:2016oda} and in the more familiar form that appears in \cite{Hashimoto:2019wct}. One minor difference between here (\ref{defE}) for $w=1$ and  \cite{Hashimoto:2019wct} is the factor of $l_2$ and the second term in parentheses inside the square root. The difference can be attributed to an additive shift in energy
\be R_y E_t  \rightarrow R_y E_t + {(1 - l_2) w \over  \lambda}~,\label{shift} \ee
which will make (\ref{defE}) recover (3.23) of \cite{Hashimoto:2019wct}. The shift in the energy is {\it{not}} ad-hoc and has its origin in the backreaction of the probe long string on the geometry. For more details on the origin of this shift, we would forward the reader to  \cite{Chakraborty:2024ugc}.

Also, from (\ref{defE}), we see that for finite $\lambda$, both positive and negative values of $w$ lead to a state with finite energy, just as we saw for the probe strings in semi-classical approximation in  (\ref{EprobeS}). In fact, we can go further in comparing (\ref{defE}) and  (\ref{EprobeS}) in the large $k$ limit. It was explained in \cite{Maldacena:2000hw} that the spectrum of long strings is bounded
\be R_y E_t^0 > {k w \over 2} \ . \ee
When this bound is saturated, the deformed spectrum is
\be R_y  E_t = {1\over 2 \lambda}(|w| - l_2 w)  \equiv R_y  E_t^{threshold} \ . \label{Eth}\ee
This is exactly what was found in (\ref{EprobeS}). In other words, we found that
\be  E_t^{threshold}  = E_t^{probe}(\rho=\infty) \ .  \ee

\subsubsection{Discrete States \label{secdisc}}

Finally, let us discuss the flow of discrete states under the deformation. In the $AdS_3$ limit, the discrete states correspond to  operator (\ref{Vjmmw}) with the condition that
\be m = \pm(j + n)~, \qquad \bar m = \pm(j + \bar n)~, \label{mmbar} \ee
and
\be {1 \over 2} \le j \le {k-1 \over 2} \ ,  \label{bound}\ee
where $n$ and $\bar n$ are non-negative integers. The signs $\pm$ correspond to the contributions from states in  ${\cal D}_j^\pm \times {\cal D}_j^\pm$ in the notation of \cite{Maldacena:2000hw}.  Condition (\ref{bound}) arises from the consideration of normalizability and spectral flow, and is part of the definition of the principle discrete series \cite{Maldacena:2000hw}.

The main difference between the continuous and the discrete states is that the Virasoro constraint constrains $j$ directly. Since the values of $j$ are bounded by (\ref{bound}), the set of allowed $h$ and  $\bar h$ are also bounded. In the $AdS_3$ limit, this bound is given in (81) of \cite{Maldacena:2000hw}.

With the $J_3 \bar J_3$ deformation, the bound deforms as well. The relevant formula is obtained by solving the Virasoro constraint (\ref{vsum}) for $j$ and imposing (\ref{bound}) but is unilluminating to write it down.

The expression for the energy spectrum obtained by solving the Virasoro constraint for $j$ and substituting it into (\ref{EK}) is very complicated, but it does simplify in the $l_2 \rightarrow 0$ limit where the geometry becomes $R_t \times \mbox{cigar}$ where $\mbox{cigar} =  SL(2,R) /U(1)$. In that limit, we recover
\be - {{\alpha' E_t}^2 \over 2}
-{2j(j - 1) \over k-2} + {m^2 \over k} + {\bar m^2 \over k}  = 0 ~,\ee
which matches (2.4) of \cite{Giveon:2016dxe}. This is one test that the spectrum we are finding is in fact correct. In the undeformed limit (\ie\  $ l_1\to 0$) however, one trivially obtains the well-known spectrum of discrete states of $AdS_3\times \mathcal{N}$ as obtained in (78) of \cite{Maldacena:2000hw}.

The general picture that emerges from these analyses is that as the deformation parameter $l_1$ is varied, some discrete states might cease to exist, whereas some new discrete states might appear. As these states appear or disappear, they are mixing with the continuous states. It is reasonable to suspect that a more intricate structure is hiding in this threshold mixing between discrete and continuum states, but we leave their exploration for future work.

\subsection{One-loop amplitude of deformed $SL(2,R) \times {\cal N}$ string theory} \label{sec4.4}

In the final subsection, we will consider the embedding of partition function $Z_{compact}^{def H_3}$ obtained in section \ref{sec:answer} into critical string theory.

\subsubsection{Spectral Decomposition of $Z_{(1,0)}^{H_3}$ \label{sec:trace}}

It is actually convenient to go back to $Z_{(1,0)}^{H_3}$ in (\ref{ZH310}).  Just like in the $c=1$ case (\ref{Zx10undef}), one expects to write $Z_{(1,0)}^{H_3}$ for the case of $H_3$ considered in \cite{Maldacena:2000kv} in the spectrally decomposed form as follows
\beq \lefteqn{
{1 \over \beta} Z_{(1,0)}^{H_3}(\tau, \bar \tau, \xi, \bar \xi, l_1)} \cr
& = & \sum_{n, \bar n, N, \bar N, w} d(N, \bar N, w)  \label{spec2} \\
&&  \left( \int_{1/2}^{(k-1)/2} dj \, e^{-2 \pi \tau_2 (\Delta_0 + \bar \Delta_0+N+\bar{N} - c/12)(j,n,\bar n, w) + 2 \pi i \tau_1 (N - \bar N - nw + \bar n w) - \beta(2j + n + \bar n)- i \beta \mu (n-\bar n)} \right.  \cr
&& \left.+  \int_0^\infty ds \int_0^1 d \alpha \, \rho(s, m, \bar m )  e^{- 2 \pi \tau_2 (\Delta_0 + \bar \Delta_0+N+\bar{N} - c/12)(s,\alpha,n, \bar n, w) + 2 \pi i \tau_1  (N - \bar N - nw + \bar n w) - \beta(2\alpha + n + \bar n) - i \beta \mu (n-\bar n)} \right) \ . \nonumber
\eeq
The last two lines in the RHS of (\ref{spec2}) correspond respectively to the contribution from the (off-shell) discrete and continuum representation states.

Here, $\rho(s,m,\bar m)$ is the expression (\ref{rhos}) which appeared earlier in the review of \cite{Maldacena:2000kv}. It is important to stress that in (\ref{spec2}), the substitution (\ref{mms}) is not imposed. This is because (\ref{mms}) is a consequence of the Virasoro constraint which is not yet imposed when writing (\ref{spec2}) for the CFT on $H_3$.

The relation (\ref{spec2}) does not appear in  \cite{Maldacena:2000kv} and to the best of our knowledge has not yet been derived by manipulating (\ref{ZH310}).\footnote{The closest thing we found in the literature is  \cite{Ashok:2020dnc} for just the discrete states.} What one can do however is to combine $Z_{(1,0)}^{def\ H_3}$ with $Z^{\cal N}$ for some CFT ${\cal N}$ and $Z^{ghost}$ and compute
\be {1 \over \beta} \Xi_{(1,0)}^{H_3} (\beta,\mu)  = \int_{\cal S} {d \tau d \bar \tau \over \tau_2^2} {1\over \beta} Z_{(1,0)}^{H_3}(\tau, \bar \tau,\beta,\mu) Z^{{\cal N}}(\tau, \bar\tau) Z^{ghost}(\tau,\bar \tau)~, \ee
for both (\ref{ZH310}) and (\ref{spec2}). The calculation using (\ref{ZH310}) was done in \cite{Maldacena:2000kv}. The fact that the calculation using  (\ref{spec2}) gives the same answer for arbitrary ${\cal N}$ is a strong constraint that makes it essentially certain that  (\ref{ZH310}) and (\ref{spec2}) are equal to each other, although it would be nice to construct a more formal proof of that statement.  We will proceed with the assumption that spectrally decomposed form (\ref{spec2}) follows from (\ref{ZH310}).

Let us proceed to compute the   ${\cal D}_j^+ \times {\cal D}_j^+$ component of $\Xi_{(1,0)}^{H_3} (\beta,\mu)$. We start by writing
\beq  {{1 \over \beta}} \Xi_{(1,0)}^{H_3\ disc} (\beta,\mu)
& = &\int_{\cal S} {d \tau_1 d \tau_2  \over \tau_2}  \sum_{n, \bar n, N, \bar N, w, h, \bar h} \int_{1/2}^{(k-1)/2} dj \ D(N, \bar N, h, \bar h, w) \\
&& \quad e^{-2 \pi \tau_2 ((\Delta_0 + \bar \Delta_0)(j,n,\bar n, w) +N+\bar{N}+ h + \bar h-2)+ 2 \pi i \tau_1 (h - \bar h  +N - \bar N - nw + \bar n w) - \beta(2j + n + \bar n) - i \beta \mu(n - \bar n)} \ . \nonumber \eeq
Just as in the $c=1$ case, the degeneracy factor $d(N, \bar N, w)$ and $D(N, \bar N, w)$ are slightly different because of the contribution of $Z^{ghost}(\tau, \bar \tau)$. We now implement the same trick as in the $c=1$ case of introducing auxiliary variables $y$ and $z$ to facilitate evaluating the $(\tau, \bar \tau)$ integrals. That is, we write
\beq  \lefteqn{{{1 \over \beta}} \Xi_{(1,0)}^{H_3\ disc} (\beta,\mu)
=\int{dy\ dz \over 2 \pi} \int_{\cal S} {d \tau_1 d \tau_2  \over \tau_2}  \sum_{n, \bar n, N, \bar N, w, h, \bar h} \int_{1/2}^{(k-1)/2} dj \ D(N, \bar N, h, \bar h,  w) e^{i z (y-j)} }\cr
&& \qquad e^{-2 \pi \tau_2 ((\Delta_0 + \bar \Delta_0)(y,n,\bar n, w) +N+\bar{N}+ h + \bar h-2)+ 2 \pi i \tau_1 (h - \bar h  +N - \bar N - nw + \bar n w) - \beta(2j + n + \bar n) - i \beta \mu(n - \bar n)} .  \eeq

At this point, the $y$ and the $(\tau, \bar \tau)$ integral can be done in closed form, leading to
\beq \lefteqn{{{1 \over \beta}} \Xi_{(1,0)}^{H_3\ disc} (\beta,\mu) = \sum_{n, \bar n, N, \bar N, w, h, \bar h} D(N, \bar N, h, \bar h,  w) } \cr
&& \left.   \int_{1/2}^{(k-1)/2} dj{1 \over i } \int {dz \over 2 \pi}\, {2 \pi \over -i z} e^{-\beta  (2 j+n+\bar n) - i \beta \mu (n - \bar n)- i z f(j)} \right|_{h - \bar h + N - \bar N - nw + \bar nw=0} ~,\eeq
where
\be f(j) =
j+\frac{(k-2) w}{2}-\frac{1}{2}  - \sqrt{\frac{1}{4}-(k-2) (h -1 + N -nw - {1 \over 2} w(w+1))} \ .
\ee
The $z$ integral then leads to
\beq \lefteqn{{{1 \over \beta}} \Xi_{(1,0)}^{H_3\ disc} (\beta,\mu)= \sum_{n, \bar n, N, \bar N, w, h, \bar h} D(N, \bar N, h, \bar h, w)} \cr
&& \left. \int_{1/2}^{(k-1)/2} dj {2 \pi  \over i }   e^{-\beta  (2 j+n+\bar n)- i \beta \mu (n - \bar n)} \Theta(f(j)) \right|_{h - \bar h  +N - \bar N - n w+ \bar nw =0} \cr
& = &
  \sum_{n, \bar n, N, \bar N, w, h, \bar h} D(N, \bar N, h, \bar h, w)  \cr
&& \qquad \left. \int_{1/2}^{(k-1)/2} dj \, { \pi  \over i }   {1 \over  \beta}  e^{-\beta (2 j+n+\bar n)- i \beta \mu (n - \bar n)}  \delta(f(j)) \right|_{h - \bar h  +N - \bar N -n w+ \bar nw =0}
\eeq
where we used integration by parts in the last line.

The function $f(j)$ is linear in $j$, and the condition that it is set to zero is equivalent to  (76) of \cite{Maldacena:2000hw} for Virasoro constraint for the discrete states in $AdS_3$. As a result, the $j$ integral in the range $1/2 \leq j \leq (k-1)/2$ will pick up the contribution of the states satisfying the Virasoro constraint and the bound on $j$, consistent with the discrete contribution of (\ref{thermalpart}). Similar analysis can be done for the states in  ${\cal D}_j^-\times {\cal D}_j^-$ sector.

The analysis for the continuous states follows a similar line of argument, where we introduce auxiliary variables $(y,z)$ and integrate in the order $y$, followed by $(\tau, \bar \tau)$, followed by $z$. This leads to an expression
\beq
{{1 \over \beta}} \Xi_{(1,0)}^{H_3\ cont} (\beta,\mu)
& =& { \pi \over i \beta }\sum_{n, \bar n, N, \bar N, w, h, \bar h} D(N, \bar N, h, \bar h, w) \cr
&& \quad \int_0^\infty ds \int_{0}^1 d \alpha \, \rho(s, m, \bar m )   e^{ -\beta(2\alpha + n + \bar n) - i \beta \mu l }
 \left.   {\partial f(\alpha) \over \partial \alpha} \delta\left(f(\alpha)\right)\right|_ {h - \bar h  +N - \bar N - nw + \bar n w = 0 }
\eeq
with
\be f(\alpha) =
 j - {1 \over 2} + \sqrt{ {1 \over 4} + (k-2) \left(h -1+ N - wn - \alpha w - {k w^2 \over 4}\right) } \ .
\ee
The condition that $f(\alpha)=0$ is equivalent to the Virasoro constraint
\be \Delta_0  + \bar \Delta_0 +N+\bar{N}+ h + \bar h - 2 = 0 \ ,\ee
with
\be m = \alpha+ n ,\qquad \bar m  = \alpha + \bar n \ . \ee

Integrating out $\alpha + n$ keeping $n - \bar n \equiv l$ fixed, we arrive at
\beq
\lefteqn{{{1 \over \beta}} \Xi_{(1,0)}^{H_3\ cont} (\beta,\mu)} \\
&=&\left.
{ \pi \over i \beta}  \sum_{l, N, \bar N, w, h, \bar h} D(N, \bar N, w)  \int_0^\infty ds  \, \rho(s, m, \bar m )  e^{ -\beta(2\alpha + n + \bar n) - i \beta \mu l }\right|_ {h - \bar h  +N - \bar N - nw + \bar n w = 0, f(\alpha) = 0 } ~,\nonumber
\eeq
which again is in agreement with (\ref{thermalpart}) including (\ref{rhos}) and (\ref{mms}).

So far we only focused on $(m,n) = (1,0)$ sector but because of the fact that we integrated over ${\cal S}$ instead of ${\cal F}$, the sum over $\Lambda$ in (\ref{foldsum}) and (\ref{foldsum2}) is already accounted for. All that is needed in order to obtain the full spacetime one-loop amplitude is to sum over $m$. Just as was the case in \cite{Polchinski:1985zf}, this sum accounts for the multi-string sectors of the string gas and does not contain any new physics in the non-interacting limit of the strings.

\subsubsection{One-loop amplitude of the $J_3 \bar J_3$ deformed $AdS_3 \times {\cal N}$} \label{secjjstemded}

Once the spectral decomposition (\ref{spec2}) of $Z_{(1,0)}^{H_3}(\tau, \bar \tau, \beta, \bar \beta)$ is understood, it is straightforward to extend the analysis for the $J_3 \bar J_3$ deformed theory using the results of section \ref{sec:answer}.  We can immediately conclude that
\beq \lefteqn{
{1 \over \beta} Z_{(1,0)}^{def\ H_3}(\tau, \bar \tau, \xi, \bar \xi, l_1)} \cr
& = & \sum_{n, \bar n, N, \bar N, w} d(N, \bar N, w)  \label{spec3} \\
&&  \left( \int_{1/2}^{(k-1)/2} dj \, e^{-2 \pi \tau_2 (\Delta+ \bar \Delta+N+\bar{N} - c/12)(j,n,\bar n, w) + 2 \pi  i \tau_1 (N - \bar N - nw + \bar n w) - \beta(2j + n + \bar n)- i \beta \mu l} \right.  \cr
&& \left.+  \int_0^\infty ds \int_0^1 d \alpha \, \rho(s, m, \bar m )  e^{- 2 \pi \tau_2 (\Delta + \bar \Delta+N+\bar{N}- c/12)(s,\alpha,n, \bar n, w) +2 \pi i \tau_1  (N - \bar N - nw + \bar n w) - \beta(2\alpha + n + \bar n) - i \beta \mu l} \right) \ , \nonumber
\eeq
where the only difference between (\ref{spec2}) and (\ref{spec3}) is that $(\Delta_0, \bar \Delta_0)$ got replaced by $(\Delta, \bar  \Delta)$.

From this, we can calculate the one-loop string amplitude of the $J_3\bar{J}_3$ deformed $AdS_3 \times {\cal N}$ theory by integrating out $(\tau, \bar \tau)$ as was done in section \ref{sec:trace}. The result of the analysis is
\beq \lefteqn{\Xi^{def}_{(1,0)}(\beta,\mu) = {1 \over \beta} \int_{{\cal S}} {d \tau_1 d \tau_2 \over \tau_2^2} Z_{(1,0)}^{def \ H_3} Z^{{\cal N}} Z^{ghost}} \cr
& =&  {1 \over \beta} \sum_{h, \bar h, N, \bar N, w}  D(h, \bar h, N, \bar N, w) \left[ \sum_{n, \bar n} e^{-\beta E_{\ttau} - i  \beta \mu l} + \int_0^\infty ds\, \rho(s) e^{-\beta E_{\ttau}(s) - i \beta \mu l} \right] ~,\label{thermalpart2}\eeq
with
\beq \rho(s) &=& -{1 \over \pi} \log \epsilon \label{rho2} \\
&&  + {1 \over 2 \pi i} {d \over ds} \log\left({\Gamma({1 \over 2} - is + \bar m -{kw \over 2}) \Gamma({1 \over 2} - is - m +{kw \over 2}) \over \Gamma({1 \over 2} + is + \bar m -{kw \over 2} ) \Gamma({1 \over 2} + is -m+{kw \over 2}) }\right) \ , \nonumber  \eeq
and
\be m(s) =  {1 \over 2}\left(E_{\ttau}(s)+l \right) =  {1 \over 2}\left( l_2 R_y E_t(s)+l \right)~, \ \ \    \bar m (s)= {1 \over 2}\left(E_{\ttau}(s)-l \right)= {1 \over 2}\left( l_2 R_y E_t(s) -l \right) \ . \label{mmbardef}\ee
The divergent piece, $\log \epsilon$, is an IR regulator whose origin can be traced back to the infinite volume of the throat region of the spacetime geometry.
$E_{\ttau}$ and $E_t$ are energies conjugate to coordinates $\ttau$ and $t$ introduced in section \ref{sec:J3J3def} and are related by  (\ref{EEtaut}). The spectrum $E_{\ttau}$ or equivalently $E_t$ and $l$ in (\ref{thermalpart2})  will solve the deformed Virasoro constraint (\ref{vsum}) following the argument of section \ref{sec:trace} and as such will automatically agree with the on-shell spectrum presented in (\ref{defE}) and (\ref{RE0}) for the continuous states and section \ref{secdisc} for the discrete states.
The density of states depends on the deformation parameter through (\ref{mmbardef}) and the Virasoro constraint.
The density of state naturally interpolates between the known results for $AdS_3 \times {\cal N}$ of \cite{Maldacena:2000kv} and $R_t \times  (SL(2,R) /U(1)) \times {\cal N}$ of \cite{Hanany:2002ev}.

As a check of the density of state (\ref{rho2}), we can compare it to the scattering phase shift of the long string states via the relation (70) of \cite{Maldacena:2000kv}
\be \rho(s) = {1 \over 2 \pi} \left( -\log\epsilon^2 + {d \delta \over ds} - {d \delta_{wall} \over ds} \right)\ . \label{rhoofs}
\ee
In order to extract the phase shift $\delta(s)$, one starts with the reflection amplitude $R(s)$ which encodes the operator relation
\be V_{\alpha,n, \bar n}(s) = R(s) V_{\alpha,n, \bar n}(-s)~, \ee
for the $SL(2,R)$ vertex operator in continuum representation. $R(s)$ encodes the asymptotic expansion of the bulk modes as is written in (64) of \cite{Maldacena:2000kv}, and an analytic expression for
\be R(s) = e^{i \delta(s)}~,\label{R(s)} \ee
and $\delta_{wall}(s)$ for $AdS_3$ is given in (79) and (80) of \cite{Maldacena:2000kv}.

In order to read off the reflection amplitude for the deformed theory, all that one needs to do is to consider the dressed vertex operator (\ref{Vdressed}),   whose reflection coefficient trivially takes the same functional form as that of  $AdS_3$ except now the quantum numbers are subject to the null gauge constraint (\ref{chargeconstraint}) and the deformed Virasoro constraints (\ref{viscons}). Imposing these constraints in the reflection coefficient (\ref{R(s)}) and substituting it in (\ref{rhoofs}) reproduces (\ref{rho2}) with (\ref{mmbardef}).

\subsubsection{Natural coordinates for the deformed $SL(2,R)$}

In this subsection, we will elaborate on the issue of choice of coordinates. It is useful to discuss this issue explicitly since there is no single canonical coordinate system that is convenient and natural to impose throughout our discussion. So far we have concentrated on coordinates $(\ttau,\phi)$ which are useful in formulating the undeformed theory. The same coordinate is not particularly natural from the target space point of view of the deformed theory. It is also important to stress that energies (\eg\ $E_\ttau$, $E_t$, \etc), temperature $\beta$, and chemical potential $\mu$, are coordinate-dependent quantities and as such it is important to keep track of the implied coordinate system when these quantities are referenced.

If one goes back to (\ref{ty0gauge}), one finds that the metric at large $\rho$, is
\be ds^2 = {k (1+\varepsilon)\over \varepsilon} \left(-{1 \over 1+\varepsilon} d \ttau^2  +d \phi^2\right) + \ldots =  {k (1+\varepsilon)\over \varepsilon}(-l_2 ^2 d \ttau^2 + d \phi^2) + \ldots~. \label{holmet}\ee
That is, the $\ttau$ coordinate gets stretched relative to the $\phi$ coordinate by a factor of $l_2$. This is precisely analogous to what we saw in (\ref{VV0ratio}) in the $c=1$ model.

The field theory dual to the undeformed and deformed $AdS_3$ ($H_3$) background is Lorentz (rotationally) invariant in the decompactified limit. Therefore from the spacetime/boundary theory point of view, a natural set of coordinates is $(\tilde \ttau, \phi)$, where
\be \tilde \ttau = l_2 \ttau ~.\label{ttau} \ee
So that the metric (\ref{holmet}) becomes
\be ds^2 = {k (1+\varepsilon)\over \varepsilon}( -d \tilde \ttau^2 + d \phi^2) + \ldots~. \ee
These are the natural dimensionless coordinates to impose the periodicity condition.  The periodicity of $\phi$ is set to $2 \pi$.
The periodicity fo $\tilde \ttau$ is
\be \tilde \beta = l_2 \beta ~. \ee
It is also natural to define $\tilde \mu$  such that
\be \tilde \mu  = {\beta \mu \over \tilde \beta}~. \ee
In these sets of coordinates and moduli parameters, a null geodesic at the boundary is $\tilde \ttau = \pm \phi$, and $\tilde \mu \tilde \beta + i \tilde \beta$ is the complex structure in the Euclidean $(\tilde x, \phi)$ coordinates where $\tilde x = -i \tilde \ttau$. Just like in (\ref{l2scale}), we can express the partition function in terms of $(\beta, \mu)$ or $(\tilde \beta, \tilde \mu)$, but unlike in (\ref{l2scale}), one can not absorb the $J_3 \bar J_3$ deformation by these rescalings.

It is also worth noting that $(\tilde \ttau, \phi)$ coordinate stays finite and regular as one interpolates between the $AdS_3$ and the  $R_t \times \mbox{cigar}$ limits.

The only issue with this coordinate system is that if we define the $J_3 \bar J_3$ deformation as we did in section \ref{sec:problem}, then $\tilde \beta = l_2 \beta$ varies with the deformation since the deformation keeps $\beta$ constant. This of course is another manifestation of (\ref{VV0ratio}). Here, $\beta$ is the periodicity of coordinate $x$ defined in (\ref{tix}), and $\tilde \beta$ is the periodicity of ``proper length'' associated with coordinate $x$. It is not difficult to go back and forth between these conventions as long as they are explicitly defined.

\section{Discussions}

In this article, we set out to compute the partition function for the $J_3 \bar J_3$ deformed compactified $H_3$,  $Z_{compact}^{def\ H_3}(\beta, \mu; \tau, \bar \tau, l_1)$, and arrived at expression (\ref{foldsum2}). The key ingredient we developed to make this computation possible is the kernel relation (\ref{H3kernel3}) and our result manifestly recover the partition function $Z_{compact}^{H_3}(\beta, \mu; \tau, \bar \tau)$ obtained in \cite{Maldacena:2000kv}  in the undeformed limit.

The sequence of steps taken in this article to compute $Z_{compact}^{def\ H_3}(\beta, \mu; \tau, \bar \tau, l_1)$ is rather long, involving the use of folding/unfolding trick, spectral decomposition of $Z_{(1,0)}^{H_3}(\beta, \mu; \tau, \bar \tau)$, and the application of the kernel formula. To better illustrate the sequence of steps, we prepared a flow chart summarizing the procedure in figure \ref{fige}. Figure \ref{fige} is the extended version of figure \ref{figb2}. In figure \ref{fige}, the application of the kernel (\ref{H3kernel3}) is the step labeled by $\Updownarrow$.

\begin{figure}
\centerline{
\begin{tabular}{ccccc}
\fbox{\parbox{2.07in}{$$Z_{compact}^{H_3}(\beta, \mu; \tau, \bar \tau)\ (\ref{ZH_3}) $$}} & $\leftrightarrow$ & \fbox{\parbox{1.85in}{$$Z_{(1,0)}^{H_3}(\beta, \mu; \tau, \bar \tau)\ (\ref{ZH310}) $$}} & $\searrow$  \\
&&&&  \fbox{\parbox{1.7in}{$$\Xi^{H_3}_{(1,0)}(\beta, \mu)\ (\ref{thermalpart}) $$}}\\
&&& $\nearrow$ \\
&& \fbox{\parbox{1.7in}{Spectral Representation  (\ref{spec2})}} \\
{\color{red}$\Bigg{\Downarrow}$} && $\Updownarrow$ \\
&& \fbox{\parbox{1.7in}{Deformed Spectral Representation  (\ref{spec3})}} \\
&& & $\searrow$ \\
&&  {\color{blue}$\Big{\updownarrow}$} &&  \fbox{\parbox{1.5in}{$$\Xi^{def}_{(1,0)}(\beta, \mu,l_1)\ (\ref{thermalpart2}) $$}}\\
\fbox{\parbox{2in}{$$Z_{compact}^{def\ H_3}(\beta, \mu; \tau, \bar \tau, l_1)\ (\ref{foldsum2}) $$}} & $\leftrightarrow$ & \fbox{\parbox{1.85in}{$$Z_{(1,0)}^{def\ H_3}(\beta, \mu; \tau, \bar \tau,l_1)\ (\ref{H3kernel3}) $$}} &
\end{tabular}
}
\caption{Flow chart for computing $Z_{compact}^{def\ H_3}(\beta, \mu; \tau, \bar \tau, l_1)$ in terms of $Z_{compact}^{H_3}(\beta, \mu; \tau, \bar \tau)$.  The relation between the leftmost column and the center column is the folding/unfolding trick. The relation between the center and the rightmost column is the embedding of $H_3$ into critical string theory $H_3 \times {\cal N}$. The double arrow $\Updownarrow$ is where the kernel (\ref{H3kernel3}) implementing the $J_3 \bar J_3$ deformation is applied. The blue arrow relating (\ref{spec3}) and (\ref{H3kernel3}) follows from the fact that the relation between (\ref{ZH310}) and (\ref{spec2}) can be generalized to the deformed case via the relation indicated as $\Updownarrow$. This figure is an extended version of figure \ref{figb2}.  The red arrow corresponds to the red arrow in figure \ref{figb2}. \label{fige}}
\end{figure}

Some of the links illustrated in figure \ref{fige} are weak. Most notable is the step relating  (\ref{ZH310})  and (\ref{spec2}). As was explained in section \ref{sec:trace}, the equivalence of (\ref{ZH310})  and (\ref{spec2}) is argued on the basis that they both lead to (\ref{thermalpart}) for arbitrary choice of ${\cal N}$.  The spectrally decomposed expression  (\ref{spec2}) is well suited to apply the kernel (\ref{H3kernel3}). The fact that the spectral decomposition of the deformed theory  (\ref{spec3}) leads to a sensibly normalized expression (\ref{thermalpart2}) for $\Xi^{def}_{(1,0)}(\beta, \mu,l_1)$ which is expressed in terms of density of states function compatible with the reflection amplitude for the long strings (\ref{rho2}) is a non-trivial consistency check that (\ref{ZH310})  and (\ref{spec2}) are equivalent. This equivalence is also used to relate  (\ref{spec3}) and (\ref{H3kernel3}). It would be more satisfying to establish the equivalence of   (\ref{ZH310})  and (\ref{spec2}) by direct computation.

One could subject the flow chart illustrated in figure \ref{fige} to more tests. For example, it should be possible to show that  (\ref{ZH310})  leads to (\ref{thermalpart2}) by embedding (\ref{ZH310}) into critical string theory on $H_3 \times {\cal N}$, acting on it by the kernel, and then evaluating the $\tau,\bar{\tau}$ integrals. We have not performed this test.

Although these additional consistency tests would be desirable, it is remarkable that explicit expressions (\ref{H3kernel3}) and (\ref{foldsum2})  can be written down.

The quantity  (\ref{foldsum2}) is the thermal partition function of $SL(2,R)$ WZW CFT deformed by exactly marginal $J_3 \bar J_3$ operator. From the $SL(2,R)$ CFT point of view, such a deformation is similar to $T \bar T$ and $J \bar T$ deformations, and a relation between the partition function of the undeformed theory and the deformed theory can be presented explicitly as an integral transform (\ref{SL2RBoltzKern}) closely resembling  (3.23) of \cite{Hashimoto:2019wct}.

What is interesting about the exactly marginal deformation of $SL(2,R)$ is that it can be treated as a component of critical string theory on deformed $AdS_3 \times {\cal N}$, which further can be interpreted as an {\it irrelevant} deformation of a boundary CFT in a holographic sense.  The one-loop vacuum amplitude with compactified time coordinate is then interpretable as the thermal partition function of the boundary theory, which in this case is two-dimensional LST.

One natural question is whether this one-loop vacuum amplitude of the deformed boundary theory can be presented in a form resembling  (3.23) of \cite{Hashimoto:2019wct}.  For this to be true implies that the integral kernel smears the spacetime moduli $(\xi, \bar \xi)$. While the deformed partition function (\ref{thermalpart2}) derived using  (\ref{SL2RBoltzKern}) does involve integral over $(\xi, \bar \xi)$, it also involves an integral over $(\tau, \bar \tau)$ which is an auxiliary parameter from the point of view of the spacetime theory. The fact that the smearing function (\ref{SL2RBoltzKern}) depends at all on $(\tau, \bar \tau)$ is an indication that the deformation is not interpretable as the strict smearing of the spacetime moduli parameter $(\xi, \bar \xi)$.

This failure for the one-loop vacuum amplitude to be like (3.23) of \cite{Hashimoto:2019wct} is also apparent from the expression for the deformed partition function  (\ref{thermalpart2}). From  (\ref{thermalpart2}) and (\ref{rho2}), we see that both the spectrum $E_\ttau(s)$ or equivalently $E_t(s)$ and the density of states $\rho(s)$ in the continuum sector are affected by the deformation. But the deformation (3.23) of \cite{Hashimoto:2019wct} only affects the spectrum.

What this suggests is that only a {\it part} of the one-loop vacuum amplitude will transform like (3.23) of \cite{Hashimoto:2019wct}. It is also easy to see which part we should isolate from the form of the expression (\ref{rho2}). In (\ref{rho2}), there are terms proportional to $\log (\epsilon)$ and terms that are independent of $\epsilon$.  For small $\epsilon$, it is the $\log(\epsilon)$ term that dominates. The $\log (\epsilon)$ term captures the physics of continuum strings living in a volume of order $\log(\epsilon)$ near the boundary of the deformed $AdS_3$ geometry. This term is such that the density of the state is independent of the deformation. The finite term in (\ref{rho2})  and the contribution from the discrete states in (\ref{thermalpart2})  are of order $\epsilon^0$ and are subleading to the $\log(\epsilon)$ term. These $\epsilon^0$ terms are interpretable as being sensitive to the physics in the core region of $AdS_3$ which is capping off the linear dilaton behavior away from the core. It then stands to reason that the $\log(\epsilon)$ term is capturing the physics of continuous strings ignoring the cap region.

What we conclude from this discussion is that
\begin{enumerate}
\item Terms proportional to $\log(\epsilon)$ in the partition function of string theory on $H_3 \times {\cal N}$ are interpretable as having a symmetric product structure, and

\item The deformation of the terms proportional to $\log \epsilon$ by $J_3 \bar J_3$ deformation behaves exactly as the single-trace $T \bar T$ deformation of a symmetric product CFT as was explained in \cite{Hashimoto:2019hqo} but includes states with zero and negative winding numbers in the string gas. The negative and zero winding number states decouple in the undeformed limit $\varepsilon \rightarrow 0$ limit, where $\varepsilon$ is the quantity defined in (\ref{varep}) parameterizing the deformation.
\end{enumerate}

It seems then that the natural scaling limit which keeps only the $\log(\epsilon)$ states and removes the discrete states is the large $\rho$ asymptotic limit of (\ref{ty0gauge}),
\beq ds^2 & = &  {k(1 + \varepsilon) \over \varepsilon} (- d \tilde \ttau^2 + d \phi^2) + {k \over 4} d \rho^2 ~, \cr
B & = &  {k  \sqrt{1 + \varepsilon}  \over \varepsilon  } d \tilde \ttau \wedge d \phi~, \label{scalinglimit} \\
 \Phi&=& -\frac{\rho}{2} +const~,\nonumber \eeq
which is nothing other than a linear dilaton background times a two-dimensional flat spacetime on a circle with some $B$-field. Here, we are using the coordinate $\tilde \ttau$ defined in (\ref{ttau}). The long string spectrum can be constructed with just the data available from string theory in the background (\ref{scalinglimit}), as explained in \cite{Chakraborty:2024ugc}.

\section*{Acknowledgements}

We thank David Kutasov for extensive conversations and feedback throughout this project.
We also thank Juan Maldacena and Jan Troost for correspondence regarding their work \cite{Maldacena:2000kv} and \cite{Ashok:2020dnc}.  The work of SC received funding under the Framework Program for Research and
“Horizon 2020” innovation under the Marie Sklodowska-Curie grant agreement number 945298. This work was supported in part by the FACCTS Program at the University of Chicago. The work of AG was supported in part by the ISF (grant number 256/22) and the BSF (grant number 2018068). The work of AH was supported in part by the U.S. Department of Energy, Office of Science, Office of High Energy Physics, under Award Number DE-SC0017647.

\appendix

\section{Kernel formula for the $SL(2,R)$ sigma model \label{sec:appa}}

In the body of this article, we derived the kernel (\ref{c1kernel3}) for the $J_x\bar J_x$ deformation of a free boson model and conjectured that the same basic kernel (\ref{SL2RBoltzKern}) can be applied to the $SL(2,R)$ model. To the extent that we reproduced the spectrum and the density of states, we can consider this conjecture passing a non-trivial check.

In this appendix, we will present a different argument leading to (\ref{SL2RBoltzKern}) by repeating the analysis of section \ref{sec:scalar} directly to the gauged $SL(2,R)$ sigma model.

We start with (\ref{basicaction}) but impose a gauge that $\bar A = \bar A_0$ and $B = B_0$ are constants.  This will lead to
\beq \lefteqn{ 2 \pi \mathcal{L}} \cr
 & =&   {k \over 2} \left( \rule{0ex}{3ex}\partial  \rho \partial \rho - \partial \theta_L (\bar \partial \theta_L + 2 l_1 \bar A_0) - (\partial \theta_R + 2 l_1 B_0) \bar \partial \theta_R - 2 (\partial \theta_R + l_1 B_0) (\bar \partial \theta_L + l_1 \bar A_0) \cosh \rho \right. \cr
 && \qquad \left. \rule{0ex}{3ex}+ (\partial \theta_R \bar \partial \theta_L-\partial \theta_L \bar \partial \theta_R) \right) \cr
&& -  ( \partial t \bar \partial t - 2l_2 B_0 \bar \partial t -2 l_2 \bar A_0  \partial t + 2 l_2^2 B_0 \bar A_0 ) \cr
&& +  \partial y \bar \partial y - 2 l_3 B_0 \bar \partial y + 2 l_3 \bar A_0   \partial y - 2 l_3^2 B_0 \bar A_0 \ ,
\eeq
along with a Faddeev-Popov determinant.  We can now integrate out $t$ and $y$ which will cancel the above-mentioned Faddeev-Popov determinant leading to
\beq 2 \pi \mathcal{L}
 & =&   {k \over 2} \left( \rule{0ex}{3ex}\partial  \rho \partial \rho - \partial \theta_L (\bar \partial \theta_L + 2 l_1 \bar A_0) - (\partial \theta_R + 2 l_1 B_0) \bar \partial \theta_R \right. \cr
&&   - 2 (\partial \theta_R + l_1 B_0) (\bar \partial \theta_L + l_1 \bar A_0 ) \cosh \rho  - 2 l_1^2 B_0 \bar A_0 \cr
&& \left. \rule{0ex}{3ex} + (\partial \theta_R \bar \partial \theta_L-\partial \theta_L \bar \partial \theta_R)  \right) \cr
&& -  4 l_2^2 B_0 \bar A_0  \ .   \label{tyintegrated}
\eeq
The first line is then recognized as the usual gauged $SL(2,R)$ WZW model \cite{Tseytlin:1992ri}. Integrating out the $SL(2,R)$ fields from the first two lines formally leads to
\beq \lefteqn{Z_{minimal}^{SL(2,R)} (\tau, \bar \tau, B_0, \bar A_0)} \cr
 &\equiv & \int [D \rho] [D\theta_L] [D \theta_R]
\exp\left[-{k \over 4 \pi } \left( \rule{0ex}{3ex}\partial  \rho \partial \rho - \partial \theta_L (\bar \partial \theta_L + 2 l_1 \bar A_0) - (\partial \theta_R + 2 l_1 B_0) \bar \partial \theta_R \right.  \right. \cr
&&  \qquad \left. \left. \rule{0ex}{3ex} - 2 (\partial \theta_R + l_1 B_0) (\bar \partial \theta_L + l_1 \bar A_0 ) \cosh \rho  - 2 l_1^2 B_0 \bar A_0  - (\partial \theta_R \bar \partial \theta_L-\partial \theta_L \bar \partial \theta_R) \right) \right] \ .
\eeq
It is then natural to expect that the relation analogous to (\ref{cftinv})
\be Z_{minimal}^{SL(2,R)} (\tau, \bar \tau, B_0, \bar A_0) = Z_{cft}^{SL(2,R)}(\tau, \bar \tau, B_0, \bar A_0)e^{\pi k \tau_2 l_1^2(B_0 - \bar A_0)^2} e^{- 4 \pi k \tau_2 B_0 \bar A_0}~, \label{mincftSL2R}\ee
would hold.  Here, $Z^{SL(2,R)}_{cft}$, like in \cite{Hashimoto:2019wct}, is referring to a quantity like (A.7) of \cite{Hashimoto:2019wct} or (\ref{spec2}) which is presented in the trace form.  The rest of the argument, including the last line of (\ref{tyintegrated}) will proceed identically as in section \ref{sec:scalar} and we will arrive at (\ref{SL2RBoltzKern}).

The statement of  (\ref{mincftSL2R}) would follow from the universality of (A.13) of \cite{Hashimoto:2019wct}, but the status of the statement of universality is presently that of a conjecture. The fact that things seem to work out for free boson and the $SL(2,R)$ model is a non-trivial test for this conjecture. It would be very nice to lift the status of this conjecture to a theorem, hopefully in the near future.


\providecommand{\href}[2]{#2}\begingroup\raggedright\endgroup

\end{document}